\documentclass[sigconf]{acmart}
\usepackage{booktabs}
\usepackage{graphicx}
\usepackage{multirow}
\newtheorem{definition}{Definition}
\AtBeginDocument{%
  \providecommand\BibTeX{{%
    \normalfont B\kern-0.5em{\scshape i\kern-0.25em b}\kern-0.8em\TeX}}}

\usepackage{booktabs}
\usepackage{multirow}
\usepackage{graphicx}
\newtheorem{theorem}{Theorem}

\newtheorem{prop}{Proposition}

\usepackage{algorithm}
\usepackage{algorithmic}

\copyrightyear{2022}
\acmYear{2022}
\setcopyright{acmcopyright}\acmConference[WWW '22]{Proceedings of the ACM Web Conference 2022}{April 25--29, 2022}{Virtual Event, Lyon, France}
\acmBooktitle{Proceedings of the ACM Web Conference 2022 (WWW '22), April 25--29, 2022, Virtual Event, Lyon, France}
\acmPrice{15.00}
\acmDOI{10.1145/3485447.3512118}
\acmISBN{978-1-4503-9096-5/22/04}

\begin{document}

\title{HRCF: Enhancing Collaborative Filtering via Hyperbolic Geometric Regularization}



\author{Menglin Yang}
\affiliation{%
  \institution{The Chinese University of Hong Kong}
      \city{Hong Kong SAR}
  \country{China}
}
\email{mlyang@cse.cuhk.edu.hk}

\author{Min Zhou}
\affiliation{%
  \institution{Huawei Noah's Ark Lab}
      \city{Shenzhen}
  \country{China}
}
\email{zhoumin27@huawei.com}

\author{Jiahong Liu}
\affiliation{%
  \institution{Harbin Institute of Technology}
    \city{Shenzhen}
  \country{China}
}
\email{jiahong.liu21@gmail.com}

\author{Defu Lian}
\affiliation{%
  \institution{University of Science and Technology of China}
    \city{Hefei}
  \country{China}
}
\email{liandefu@ustc.edu.cn}

\author{Irwin King}
\affiliation{%
  \institution{The Chinese University of Hong Kong}
    \city{Hong Kong SAR}
  \country{China}
}
\email{king@cse.cuhk.edu.hk}

\renewcommand{\shortauthors}{Yang, et al.}

\begin{abstract}
In large-scale recommender systems, the user-item networks are generally scale-free or expand exponentially. The latent features (also known as embeddings) used to describe the user and item are determined by how well the embedding space fits the data distribution.
Hyperbolic space offers a spacious room to learn embeddings with its negative curvature and metric properties, which can well fit data with tree-like structures. Recently, several hyperbolic approaches have been proposed to learn high-quality representations for the users and items.
However, most of them concentrate on developing the hyperbolic similitude by designing appropriate projection operations, whereas many advantageous and exciting geometric properties of hyperbolic space have not been explicitly explored. 
For example, one of the most notable properties of hyperbolic space is that its capacity space increases exponentially with the radius, which indicates the area far away from the hyperbolic origin is much more embeddable. Regarding the geometric properties of hyperbolic space, we bring up a \textit{Hyperbolic Regularization powered Collaborative Filtering} (HRCF)  and design a geometric-aware hyperbolic regularizer. Specifically, the proposal boosts optimization procedure via the root alignment and origin-aware penalty, which is simple yet impressively effective. Through theoretical analysis, we further show that our proposal is able to tackle the over-smoothing problem caused by hyperbolic aggregation and also brings the models a better discriminative ability. We conduct extensive empirical analysis, comparing our proposal against a large set of baselines on several public benchmarks. The empirical results show that our approach achieves highly competitive performance and surpasses both the leading Euclidean and hyperbolic baselines by considerable margins. Further analysis verifies the rationality and effectiveness of the proposal for robust, deeper, and lightweight neural graph collaborative filtering.
\end{abstract}

\maketitle

\section{Introduction}
The advent of modern information technology is associated with the excessive quantity of daily information and data, which makes it difficult to understand an issue or make decisions effectively. To alleviate information overload, recommender systems~\cite{ma2008sorec,ma2009learning,zheng2009wsrec,zhang2019star,lian2020personalized,chen2021towards}, which seek to predict the preference of a user would give to an item by capturing the item's characteristics and the user's behaviors, have been put forward and widely applied.
As one of the most widely used techniques in personalized recommendation, collaborative filtering comes from the fact that similar users would exhibit a similar preference for items.  

\begin{figure}[!t]
\vspace{10pt}
\centering
\includegraphics[width=0.32\textwidth]{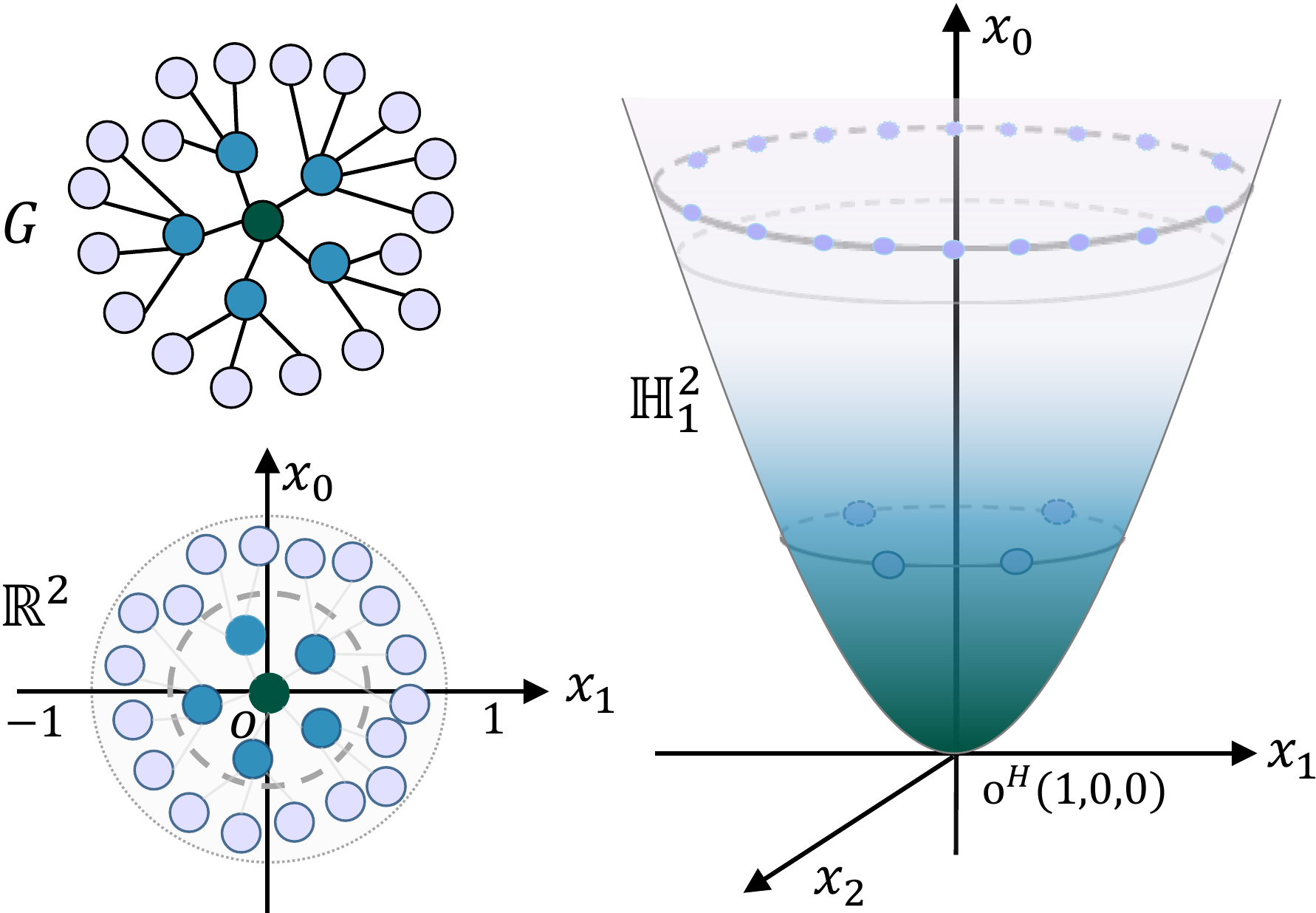}
\caption{Illustration of a tree-like graph/network $G$, Euclidean $\mathbb{R}^2$ and hyperbolic space $\mathbb{H}^2$ with the same radius 1. Many real-world networks, e.g., user-item networks, knowledge graphs are tree-like as shown in Figure~\ref{fig:degree} and can be re-organized as the $G$. Encoding $G$ in Euclidean space $\mathbb{R}^2$ and hyperbolic space $\mathbb{H}^2$ results in quite different results.
As we can see hyperbolic space has a larger room to make nodes more separable and distinguishable while the Euclidean space is too narrow to embed $G$ well. 
} 
\vspace{-10pt}
\label{fig:euclidean_hyperbolic}
\end{figure}

\begin{figure*}[t]
\centering
\includegraphics[width=4.3cm]{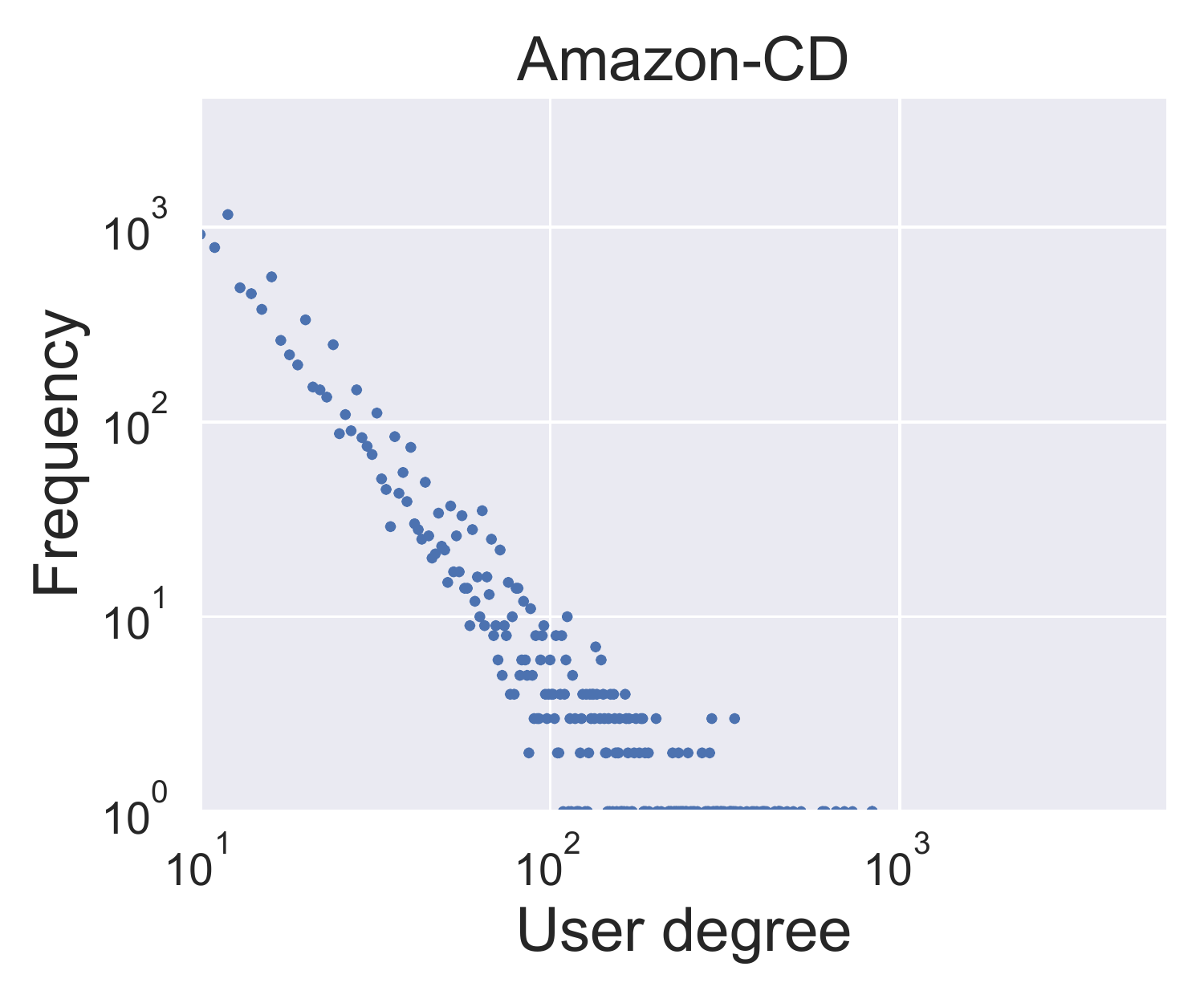}
\includegraphics[width=4.3cm]{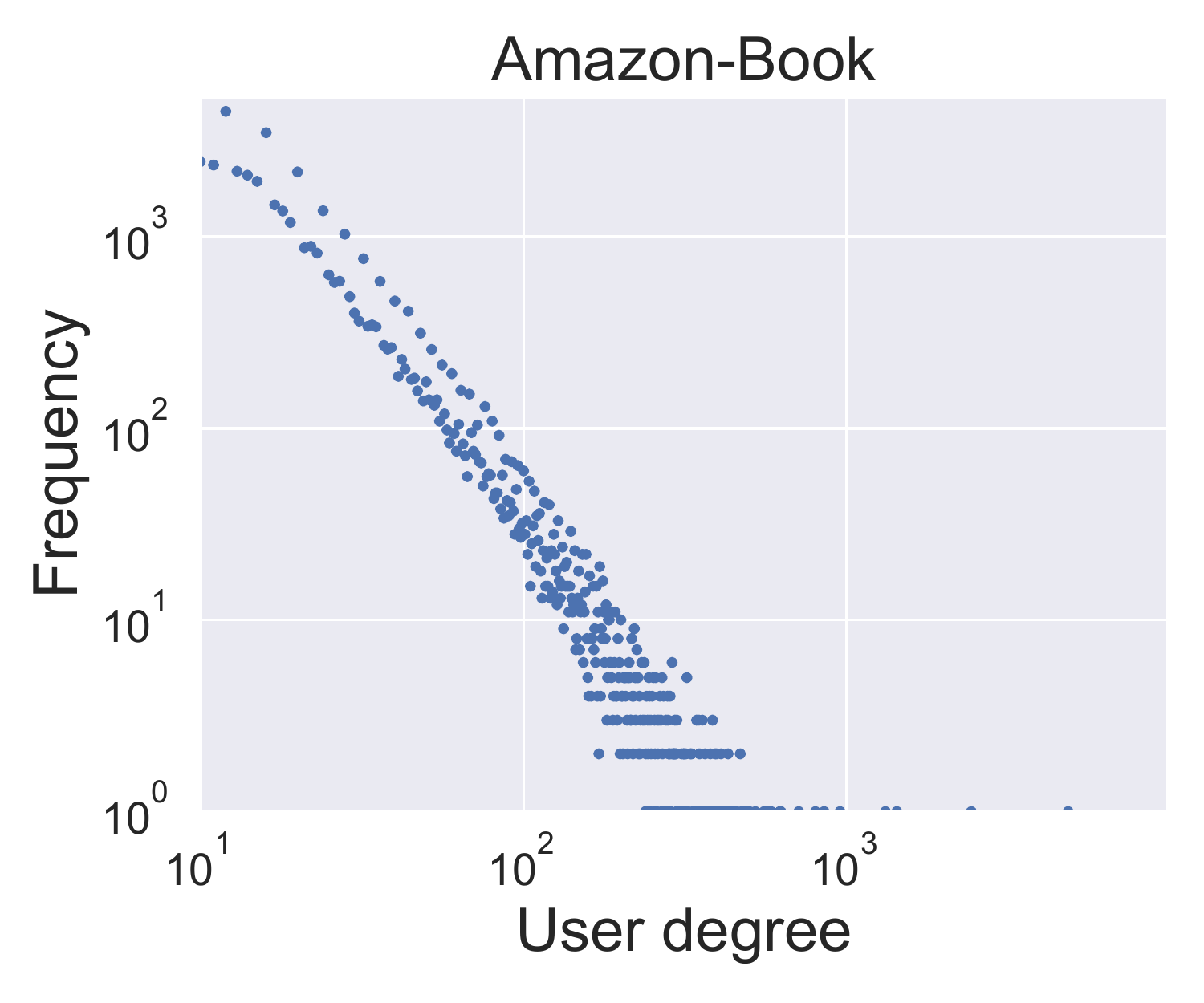}
\includegraphics[width=4.3cm]{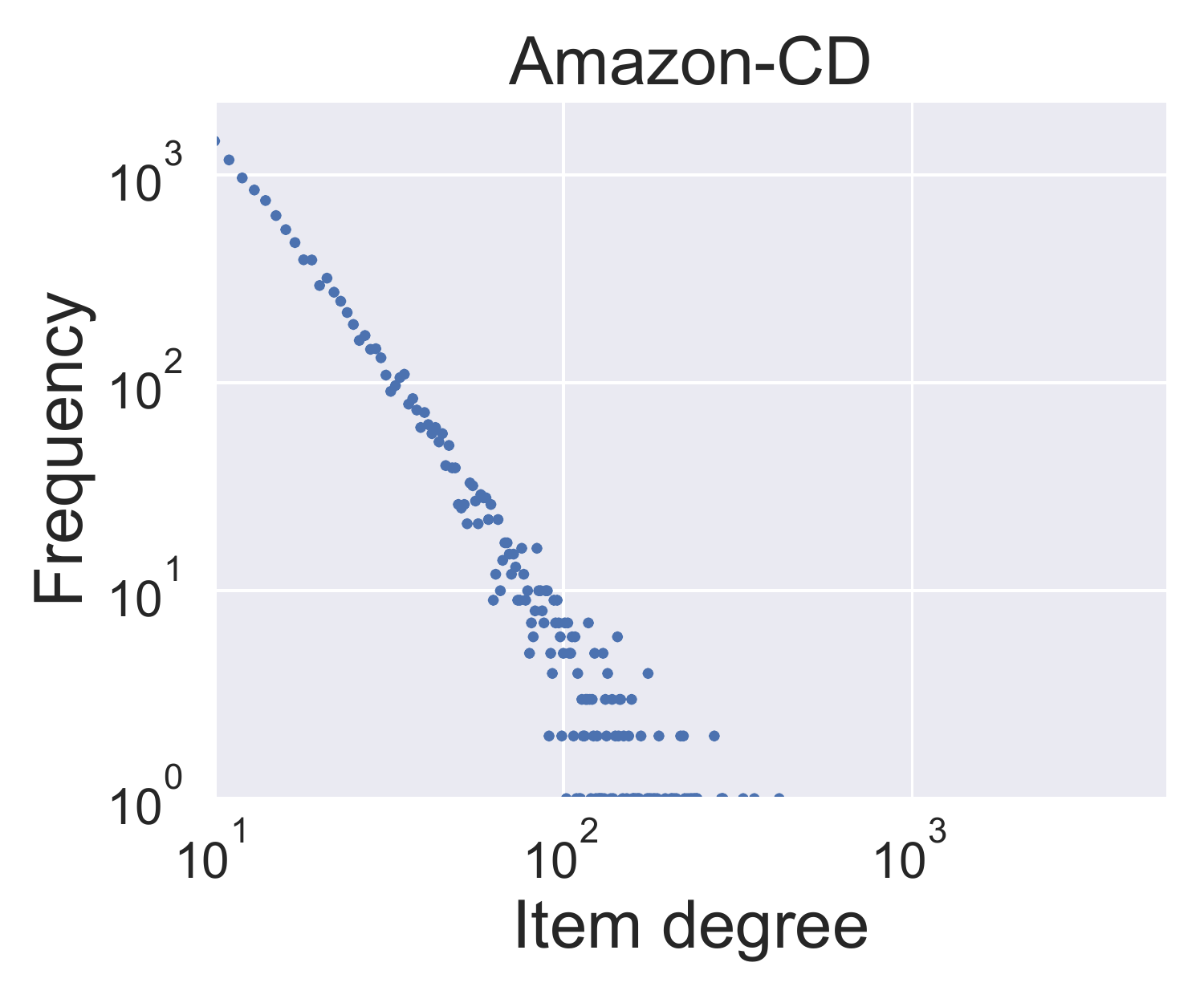}
\includegraphics[width=4.3cm]{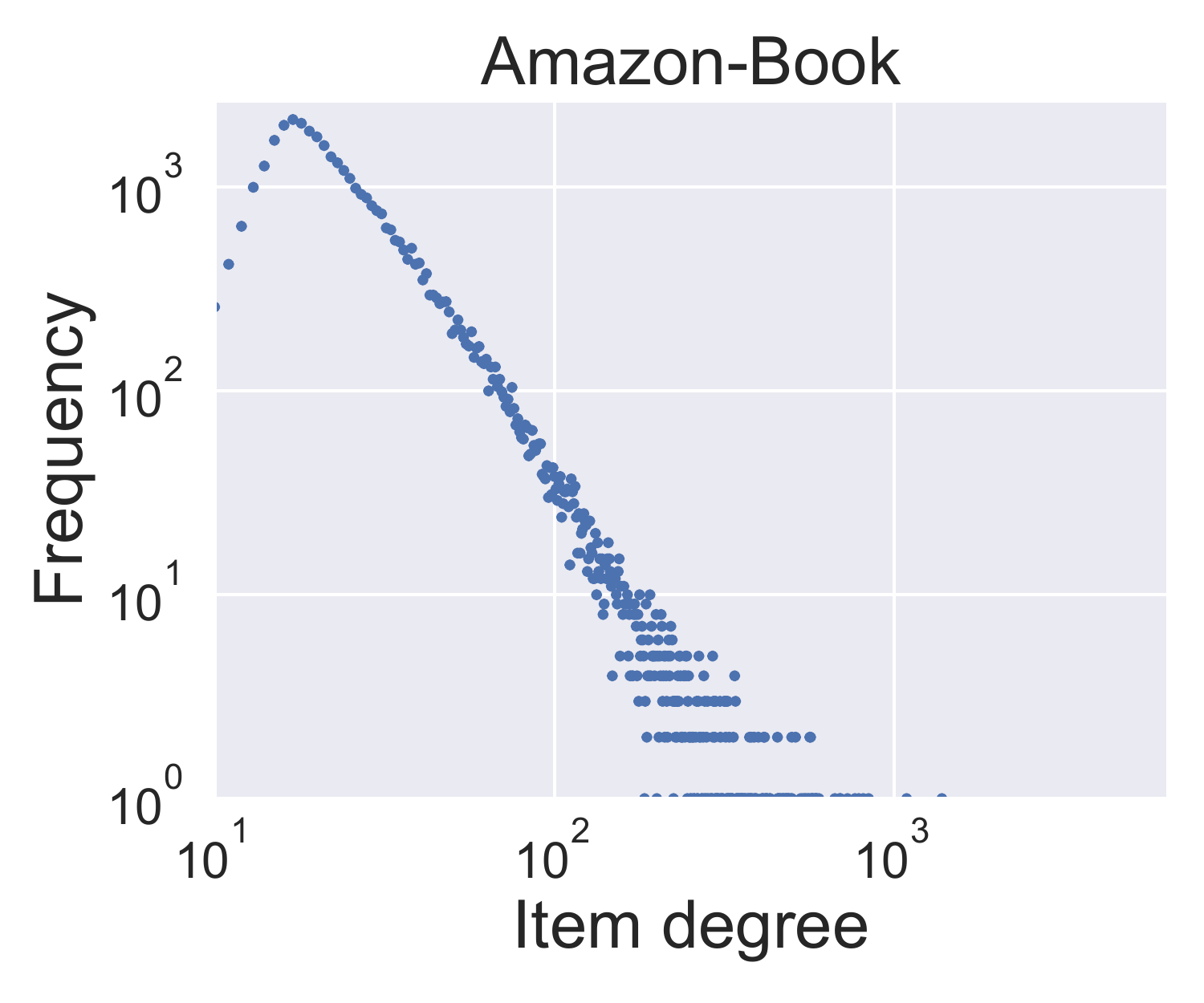}
\caption{Degree distributions of users and items in Amazon-CD and Amazon-Book datasets. The user (out-)degree represents the preference of  the user and the item (in-)degree denotes how many users like it. All of them are asymptotically power-law distributed, which indicates that the unpopular items and inactive users account for the majority while the popular and active are the minority. Note that we discard the unrepresentative node when its degree is less than 10 when plotting the above figures.}
\label{fig:degree}
\vspace{-10pt}
\end{figure*}

The most common collaborative filtering paradigm is to project both users and items into low-dimensional latent space and utilize the historical interactions to predict user's preferences. The early methods are mainly based on matrix factorization~\cite{koren2009matrix,koren2008factorization,VAECF2018,nmf_cf}. Recent works suggest explicitly integrating the high-order collaborative signal by formulating the user-item as a bipartite graph, in which vertices represent users or items, and edges denote their interactions. Then, the graph neural network is applied to extract the high-order relationships between users and items via the message propagation paradigm ~\cite{wang2019ngcf,he2020lightgcn,chen2021attentive}. 

Though the graph-based methods have made impressive achievements, most of them are built in Euclidean space and are restricted by the following two limitations.  
\textit{First}, as analyzed by~\citeauthor{NIPS2014_556f3919}~\cite{NIPS2014_556f3919}, the representation ability of Euclidean space for complex patterns is inherently bounded by its dimensionality. In other words, the volume of a  low-dimensional Euclidean based latent space is too small to well accommodate the large number of users and items of a real-world recommendation network.  
\textit{Second}, the Euclidean algorithms overlook the inherent structure of a user-item network, causing information loss.
In particular, the preference of the user and the popularity of the item are found to generally follow the power-law distribution as shown in Figure~\ref{fig:degree}, which can be traced back to the tree-like structure~\cite{ravasz2003hierarchical}. Taking the items of books as an example, the sales of bestsellers may be million-level, while the sales of general books are around thousands or less per year. It is worth mentioning that the number of general or unpopular books account for the majority. Similarly, the user's preferences also present such a power-law distribution. As pointed out by Bourgain’s theorem~\cite{linial1995geometry}, Euclidean space even with the unbounded number of dimensions still fails to obtain comparably low distortion for the data with tree-like latent anatomy.

Different from the Euclidean space which is flat, the hyperbolic space is a curved space that can be regarded as a continuous tree. As shown in Figure~\ref{fig:euclidean_hyperbolic}, the volume of Euclidean space expands polynomially while a hyperbolic space grows exponentially with its radius. Hence, it gains natural advantages in abstracting scale-free data.
In addition, \citeauthor{khrulkov2020hyperbolic}~\cite{khrulkov2020hyperbolic} found that the samples placed close to the origin are ambiguous and hard to be correctly distinguished compared with those near the boundary. In the other words, if a large number of inactive users and unwelcome items could be optimized to be away from the origin of hyperbolic space, or equivalently, to approach the boundary, the learned embeddings are then more separable and descriptive.

Based on the above observations, in this work, we propose a novel and effective hyperbolic regularized collaborative filtering, HRCF, for superior recommendation performance. HRCF is designed to take full advantage of the geometric properties of hyperbolic space via aligning the root and adding the origin-aware penalty, pushing the overall embeddings far away from the hyperbolic origin which (1) enhances the latent hierarchies formation and (2) takes full advantages of embedding space.  
Moreover, with the hyperbolic geometric regularization, the node distance over all node pairs would be enlarged by optimizing far from the origin according to our theoretical analysis, which helps to well handle the feature shrinking or oversmoothing problem caused by the hyperbolic graph aggregation.
We conduct a comprehensive empirical analysis and compare HRCF against a large set of baselines. Our approach achieves highly competitive results and outperforms leading graph-based hyperbolic and Euclidean models. We further study the properties of the learned hyperbolic embeddings and show that it offers meaningful insights into robust, deeper, and lightweight neural graph collaborative filtering. To summary, the contributions of our work are as follows:
\begin{itemize}
    \item  We propose a simple yet effective geometric regularization-based hyperbolic collaborative filtering, HRCF, which takes full advantage of the geometric properties. To the best of our knowledge, this is the first work that takes a geometric view into consideration for the hyperbolic recommender system.
    
    \item We theoretically justify that the hyperbolic aggregation and our regularizer rigorously follow the Lorentz constraints, which guarantees that the operation will not lead to a deviation of the hyperbolic sheet and ensures the correctness of the follow-up operations.
    
    \item We present a theoretical analysis that the proposed HRCF can also tackle the oversmoothing problem caused by the hyperbolic graph convolution and enhance the recommendation performance as well.
    
    \item Extensive experiments demonstrate the effectiveness of our proposal as HRCF refreshes the records of hyperbolic recommender systems on various benchmark datasets and retains the exciting performance with smaller embedding sizes, fewer training epochs, and deeper graph convolutional layers.
\end{itemize}

\section{Related work}
\subsection{Graph Neural Networks}
In the field of graph representation learning, graph neural networks have recently received much interest recently~\cite{gcn2017,GAT,song2021graph, song2021semi,FeatureNorm2020,li2020autograph,fu2020magnn,zhang2019star,zhang2018gaan}. In the recommender system, the user-item interaction data can be represented by a bipartite graph, where the link represents the connection between the corresponding user and item.  Graph neural networks are able to explicitly encode the crucial collaborative signal of user-item interactions to enhance the user/item representations through the propagation process. There have been two distinct families of GNNs proposed, namely spectral approaches and spatial methods.. Spectral methods~\cite{bruna2013spectral,gcn2017,wu2019simplifying} mainly contain a series of graph filters, which rely on eigendecomposition of  the graph Laplacian matrix. 
Spatial methods~\cite{GAT,hamilton2017SAGE} perform feature aggregation directly by the local neighborhood.
In recommendation fields, NGCF~\cite{wang2019ngcf} generalizes GNN into the field of collaborative filtering. LightGCN~\cite{he2020lightgcn} further shows that the feature transformation and non-linear activation bring little effect and can be removed for collaborative filtering. Nonetheless, existing graph neural networks are mainly built in Euclidean space, which may underestimate the implicit distribution of the user-item network. Recent studies~\cite{hgcn2019,liu2019HGNN,lgcn} show that the hyperbolic space is more embeddable especially when the graph-structured data shows a hierarchical and scale-free distribution.

\subsection{Hyperbolic Graph Neural Networks}

Hyperbolic geometry has attracted increasing attention in network science communities. Recent works~\cite{liu2019HGNN,hgcn2019,zhang2021hyperbolic,lgcn,liu2022enhancing} generalize the graph neural networks into hyperbolic space. HGNN~\cite{liu2019HGNN}, HGCN~\cite{hgcn2019}, and HGAT~\cite{zhang2021hyperbolic} achieve graph convolutions at the tangent space while LGCN~\cite{lgcn} achieves graph convolution rigorously in the hyperbolic manifold. HGNN~\cite{liu2019HGNN} focuses more on graph classification tasks and provides a simple extension to dynamic graph embeddings. HGAT~\cite{zhang2021hyperbolic} introduces a hyperbolic attention-based graph convolution using algebraic formalism in gyrovector and focuses on node classification and clustering tasks. HGCN~\cite{hgcn2019} introduces a local aggregation scheme and develops a learnable curvature method for hyperbolic graph learning. LGCN~\cite{lgcn} aggregates the neighborhood information by the centroid of Lorentzian distance. HGCL~\cite{liu2022enhancing} recommends utilizing contrastive learning to improve the hyperbolic graph learning further. Besides, works in~\cite{gu2019learning,zhu2020gil} propose to learn representations over mixed spaces. 

\subsection{Hyperbolic Recommender Systems}
There are few studies that have started to leverage the power of hyperbolic space for recommender systems up to now. HyperML~\cite{HyperML2020} generalizes metric learning into hyperbolic space by additionally considering the distortion. HSCML~\cite{zhang2021we} proposes a hyperbolic social
collaborative metric learning by pushing the socially related users close to each other. HyperSoRc~\cite{wang2021hypersorec} exploits the hyperbolic user and item presentations with multiple social relations. HME~\cite{feng2020hme} studies the joint interactions of multi-view information in hyperbolic space for next-poi recommendation. HGCF~\cite{sun2021hgcf} captures higher-order information in user-item interactions by incorporating multiple levels of neighborhood aggregation through a tangential hyperbolic GCN module. LKGR~\cite{chen2021modeling} presents a knowledge-aware attention mechanism for hyperbolic recommender systems. Most of them try to generalize existing Euclidean models to hyperbolic space while the inherent advantages of hyperbolic space are seldom well considered.

\section{PRELIMINARIES}
Hyperbolic geometry is a non-Euclidean geometry with a constant negative curvature. Hyperboloid manifold ($\mathbb{H}_K^n$, \textit{c.f.}~Definition~\ref{def:lorentz_model}), one of the typical hyperbolic models, has widely applied in recent works~\cite{nickel2018learning,hgcn2019,liu2019HGNN,lgcn}. For any $\mathbf{x}\in \mathbb{H}_K^n$, there is a tangent space $\mathcal{T}_\mathbf{x}\mathbb{H}_K^n$ around $\mathbf{x}$ approximating $\mathbb{H}_K^n$, which is an $n$-dimensional vector space (\textit{c.f.}~Definition \ref{def:lorentz_tangent_space}). To realize the projection between hyperbolic space $\mathbb{H}_K^n$ and $\mathcal{T}_\mathbf{x}\mathbb{H}_K^n$, we can resort to the exponential and logarithmic map (\textit{c.f.} Definition~\ref{def:lorentz_exponential_map}). The hyperbolic original point of hyperboloid manifold $\mathbf{o}:=\{\sqrt{K}, 0, \cdots, 0\} \in \mathbb{H}^n_K$ is a common choice as the reference point to perform exponential and logarithmic operations.

\begin{definition}[Hyperboloid manifold]
\label{def:lorentz_model}
An $n$-dimensional hyperboloid manifold (also called Lorentz model) with the negative curvature $-1/K (K>0)$ is defined as the Riemannian manifold $(\mathbb{H}^n_K, g_\mathcal{L})$, where $\mathbb{H}^n_K =\{\mathbf{x}\in\mathbb{R}^{n+1}:\langle\mathbf{x},\mathbf{x}\rangle_\mathcal{L}=-K, x_0>0\}$, $g_\mathcal{L}=\eta$ $(\eta = \mathbf{I}_n$ except $\eta_{0,0}=-1)$ and $\langle\cdot , \cdot\rangle_\mathcal{L}$ is the Lorentz inner product given by Definition~\ref{def:lorentz_inner_product}. 
\end{definition}

\begin{definition}[Lorentz Inner Product]
\label{def:lorentz_inner_product}
Let $\mathbf{x,y}\in \mathbb{H}^n_K$, then the Lorentz inner product is defined as:
\begin{equation}
    \langle\mathbf{x},\mathbf{y}\rangle_\mathcal{L}:=-x_0y_0 + \sum_{i=1}^n x_iy_i.
\label{equ:inner_product}
\end{equation}
\end{definition}

\begin{definition}[Tangent Space] 
\label{def:lorentz_tangent_space}
The {tangent space} $\mathcal{T}_\mathbf{x}\mathbb{H}_K^n$ $(\mathbf{x}\in \mathbb{H}_K^n)$ is defined as the first-order approximation of $\mathbb{H}_K^n$ around $\mathbf{x}$:
\begin{equation}
    \mathcal{T}_\mathbf{x}\mathbb{H}_K^n:=\{\mathbf{v}\in \mathbb{R}^{n+1}: \langle\mathbf{v},\mathbf{x}\rangle_\mathcal{L} = 0\}.
\end{equation}
\end{definition}

\begin{definition}[Exponential \& Logarithmic Maps]
\label{def:lorentz_exponential_map}
For $\mathbf{x}\in \mathbb{H}_K^n$, $\mathbf{y}\in \mathbb{H}_K^n$ and $\mathbf{v}\in\mathcal{T}_\mathbf{x}\mathbb{H}_K^n$ such that $\mathbf{v} \neq \mathbf{0}$ and $\mathbf{y} \neq \mathbf{x}$, there exists a unique geodesic $\gamma:[0,1]\to\mathbb{H}_K^n$ where $\gamma(0)=\mathbf{x}, \gamma^\prime(0)=\mathbf{v}$.
The exponential map $\exp_\mathbf{x}: \mathcal{T}_\mathbf{x}\mathbb{H}_K^n \to \mathbb{H}_K^n$ is defined as $\exp_{\mathbf{x}}(\mathbf{v})=\gamma(1)$. Mathematically, 
\begin{equation}
    \exp_{\mathbf{x}}^{K}(\mathbf{v})=\cosh \left(\frac{\|\mathbf{v}\|_{\mathcal{L}}}{\sqrt{K}}\right) \mathbf{x} + \sqrt{K} \sinh\left(\frac{\|\mathbf{v}\|_\mathcal{L}}{\sqrt{K}}\right){\frac{\mathbf{v}}{\|\mathbf{v}||_{\mathcal{L}}}},
\end{equation}
where $\|\mathbf{v}\|_\mathcal{L} = \sqrt{\langle \mathbf{v}, \mathbf{v}\rangle _\mathcal{L}}$ is the Lorentzian norm of $\mathbf{v}$.
The logarithmic map $\log_\mathbf{x}$ is the inverse of exponential $\exp_\mathbf{x}$, which is given by
\begin{equation}
    \log_{\mathbf{x}}^{K}(\mathbf{y})=d_{\mathcal{L}}^K(\mathbf{x},\mathbf{y})\frac{\mathbf{y}+\frac{1}{K}\langle \mathbf{x}, \mathbf{y} \rangle_\mathcal{L}\mathbf{x}}{\|\mathbf{y} + \frac{1}{K}\langle \mathbf{x}, \mathbf{y} \rangle_\mathcal{L}\mathbf{x}\|_{\mathcal{L}}},
\end{equation}
where $d_\mathcal{L}^K(\cdot, \cdot)$ is the distance between two points $\mathbf{x}, \mathbf{y}\in \mathbb{H}^n_K$, which is formulated as:
\begin{equation}
    d_\mathcal{L}^K(\mathbf{x}, \mathbf{y}) = \sqrt{K}\mbox{arcosh}(-\langle \mathbf{x}, \mathbf{y}\rangle _\mathcal{L}/K).
\end{equation}
\end{definition}

In this work, we fix the $K$ as 1 for simplicity, i.e., the curvature is $-1$. For brevity, we omit the $K$ in the following sections.

\section{Methodology}
In this section, we first introduce the typical framework of (hyperbolic) graph neural collaborative filtering. On this basis, we further elaborate our method, HRCF.

\subsection{(Hyperbolic) Graph Neural Collaborative Filtering}
The basic idea of graph neural collaborative filtering~\cite{he2020lightgcn,wang2019ngcf,sun2021hgcf,chen2021modeling} is to learn representation for nodes by extracting the high-order interactions via the message aggregation. 
Similar to Euclidean graph neural collaborative filtering, there are three components in hyperbolic settings: (1) hyperbolic embedding initializing layer; (2) hyperbolic message aggregation; (3) hyperbolic prediction layer. In the following, we will present the details.

\textbf{Hyperbolic embedding initializing layer} is to generate hyperbolic initial embeddings for users and items. Similar to Gaussian-based initialization in the Euclidean space, a hyperbolic Gaussian sampling is applied in hyperbolic recommender systems~\cite{sun2021hgcf,wang2021hypersorec,chen2021modeling}. In particular, an initial hyperbolic node (including the user and item) state $\mathbf{x}^{H,0}\in\mathbb{H}^n$ is given by
\begin{equation}
    \textbf{x}^{H,0} = \exp_\mathbf{o}(\mathbf{x}^{\mathcal{T},0}),
    \label{equ: initialization}
\end{equation}
where $\mathbf{x}^{\mathcal{T},0} = (0,\mathbf{x}^{E})$ and $\mathbf{x}^{E}\in\mathbb{R}^n$ is sampled from multivariate Gaussian distribution, the superscript $0$ indicates the initial or first layer state, $H/\mathcal{T}/E$ denote the state in hyperbolic/tangent/Euclidean space. Unless otherwise stated, the tangent space of this work denotes that it is located at origin.

\textbf{Hyperbolic message aggregation} generally consists of three components~\cite{hgcn2019,liu2019HGNN}: linear transformation, neighborhood aggregation, and non-linear activation. Recent studies~\cite{he2020lightgcn} have shown that the linear transformation brings no benefit and even degrades model effectiveness since there are no meaningful initial attributes in the setting of collaborative filtering. Moreover, non-linear activation increases the computation burden and can lead to significant over-fitting on highly sparse user-item datasets. Thus, both of them are ignored in previous studies, including Euclidean space~\cite{he2020lightgcn} and hyperbolic space~\cite{sun2021hgcf} models. 

The hyperbolic neighborhood aggregation is computed by aggregating neighborhood representations of user and item from the previous aggregation. More specifically, the hyperbolic initial state $\mathbf{x}^{H,0}\in\mathbb{H}^n$ is first projected to the tangent space $\mathcal{T}_\mathbf{o}\mathbb{H}^n$ by logarithmic map $\log_\mathbf{o}(\cdot)$, which is given by\footnote{Note that this step can be removed by using $\mathbf{x}^{\mathcal{T},0}$ directly since $\mathbf{z}=\log_\mathbf{o}(\exp_\mathbf{o}(\mathbf{z}))$.}
\begin{equation}
    \mathbf{x}^{\mathcal{T},0} = \log_\mathbf{o}(\mathbf{x}^{H, 0}).
\end{equation}
Then, the $\ell$-th aggregation for the embedding of user $u$ and item $i$ is:
\begin{equation}
\label{equ:neighbors'aggregation_u}
    \mathbf{x}_u^{\mathcal{T},\ell+1} = \sum_{i\in \mathcal{N}_u}\frac{1}{|\mathcal{N}_u|}\mathbf{x}_i^{\mathcal{T},\ell},
\end{equation}
\begin{equation}
\label{equ:neighbors'aggregation_i}
    \mathbf{x}_i^{\mathcal{T},\ell+1} = \sum_{u\in \mathcal{N}_i}\frac{1}{|\mathcal{N}_i|}\mathbf{x}_u^{\mathcal{T},\ell},
\end{equation}
where $|\mathcal{N}_u|$ and $|\mathcal{N}_i|$ are the number of one-hop neighbors of $u$ and $i$, respectively.

These $\ell$ hidden states are then aggregated with sum-pooling (including both $i$ and $u$) for multi-order information extraction:
\begin{equation}
    \mathbf{x}^{\mathcal{T},\text{sum}} = \sum_{\ell}(\mathbf{x}^{\mathcal{T},1}, \cdots, \mathbf{x}^{\mathcal{T},\ell}).
    \label{equ:multiple aggregation}
\end{equation}
Finally, it is projected back to the hyperbolic space with the exponential map,
\begin{equation}
    \mathbf{e}^H=\exp_\mathbf{o}(\mathbf{x}^{\mathcal{T},\text{sum}}).
\end{equation}

\textbf{Hyperbolic prediction layer and loss function}.
With hyperbolic message aggregation, the embedding of a user or an item encodes rich structure information. 
To predict the preference of a user for an item, the hyperbolic distance $d_\mathcal{L}$ is utilized for the prediction. Given that we are considering the rank of favored items, the negative form can be utilized to make predictions., i.e, $p(u,i)=-{d_\mathcal{L}^2(\mathbf{e}_u^H, \mathbf{e}_i^H)}$. The intuitive understanding is that when item $i$ is closer to user $u$ in the embedding space, $i$ is more preferred by $u$.
To train the embedding of user $\mathbf{e}_u$ and item $\mathbf{e}_{i/j}$, hyperbolic margin ranking loss is used for training~\cite{HyperML2020,sun2021hgcf}.
\begin{equation}
    L_{\text{margin}}(u,i,j) = \max(d_\mathcal{L}^2(\mathbf{e}_u^H, \mathbf{e}_i^H)- d_\mathcal{L}^2(\mathbf{e}_u^H, \mathbf{e}_j^H) + m, 0),
    \label{equ: ranking_loss}
\end{equation}
where the subscript $(u, i)$ represents the positive user-item pair, $(u, j)$ denotes the negative pair, and $m$ is the margin. The hyperbolic margin ranking loss is based on the hyperbolic distance $d_{\mathcal{L}}$, increasing the proximity of the positive user-item pair and limiting the connectivity of the negative user-item pair.

\subsection{Hyperbolic Regularized Collaborative Filtering (HRCF)}
The recent endeavors generalize the existing Euclidean neural graph collaborative filtering to  hyperbolic space and obtain great success. However, these pioneering works are the preliminary attempts, bringing and connecting the hyperbolic geometry to the field, and many  properties of this cheerful space have not been well exploited.

One of the remarkable properties of hyperbolic geometry is that its capacity increases exponentially with the radius. As shown in Figure~\ref{fig:euclidean_hyperbolic}, a hyperbolic space offers a larger embedding space compared with the Euclidean one. Besides, the area far away from the origin is much roomier than that close to the origin. Moreover, as empirically observed by~\cite{khrulkov2020hyperbolic}, the learned embeddings located near the boundary\footnote{The model in~\cite{khrulkov2020hyperbolic} is Poincar\'e model and the boundary is far away from the origin.} are easier to classify. Intuitively, the samples are more distinguishable if we enforce the embeddings to move towards the boundary.  However, simply pushing the overall embeddings far away from the origin may destroy the internal structure (i.e., long-tail distribution) since the embeddings may not align well in the space at the beginning. 
For instance, the number of popular items accounts for much fewer and it is better to embed them closer to the origin or higher level, and arrange the less popular item more distant from the origin which accounts for the majority. To get the favorable (i.e., more robust and distinguishable) embeddings, we need to address the following two challenges: \textit{(1) How to facilitate the model to preserve and efficiently formulate the intrinsic structure; (2) How to encourage the overall embeddings to move away from the origin.}

Our basic idea is \textit{(1) to find the root node and align it with the hyperbolic origin at first.} In this way, the root node is always placed at the highest level. Then, other nodes would also be arranged well by the loss function~(Equation~(\ref{equ: ranking_loss})). \textit{(2) to further encourage the overall nodes to be of a large norm.}  Thereby,  the final embeddings can well preserve long-tail distribution and take full advantage of the hyperbolic space. 
However, how to identify the root node is not trivial.  The hierarchy information is usually not given explicitly, and multiple root nodes may exist in a real-world user-item network. For example, there are many genres of books (e.g., detective, fantasy, fiction, and romance), and each of them has the most popularity. Aligning one of them with the origin will hinder the structure of other genres. In this work, we identify the root node from the feature aspect that is the embedding center (or midpoint), which is given by
\begin{equation}
\mathbf{x}_{\text{root}}^\mathcal{T} = \frac{1}{N}\sum_{i=1}^{N}\mathbf{x}^{\mathcal{T},\text{sum}}_i,
\label{equ: embedding root}
\end{equation}
where $N$ is the total number of users and items. Mathematically, the embedding center $\mathbf{x}_{\text{root}}^\mathcal{T}$ can be regarded as the centroid, which minimizes the sum of squared tangential distances between itself and each point in the set, which shows the similar property of a root node in a tree. In addition, it has \textit{three advantages} by defining a root node in this way. \textit{First of all}, it is easy to compute as it only costs $O(n)$ time complexity and it  can be further accelerated by the parallel computing, which is applicable when dealing with large-scale datasets; \textit{Second}, it can handle cases with single or multiple roots since the midpoint is always a unique one; \textit{Last}, this defined root is task-relevant which can be adjusted adaptively according to the optimization target. 

Then we make root alignment with the hyperbolic origin,
\begin{equation}
    \bar{\mathbf{x}}_i^{\mathcal{T},\text{sum}} = {\mathbf{x}}_i^{\mathcal{T}, \text{sum}} - \mathbf{x}^\mathcal{T}_{\text{root}}.
    \label{equ:alignment}
\end{equation}
After that, the second problem can be solved by adding a designed regularizer, which prevents the norm of each item (or/and user) from being too small in the loss function. The mathematical formulation is given as:
\begin{equation}
    \begin{aligned}
        &\bar{\mathbf{x}}_\text{norm} = \frac{1}{N}\sum_{i=1}^N \|\bar{\mathbf{x}}_i^{\mathcal{T},\text{sum}}\|_2^2, \\
        &L_{\text{reg}}  = \frac{1}{(\bar{\mathbf{x}}_\text{norm})^{1/2}}.
    \end{aligned}
    \label{equ:x_norm}
\end{equation}
Totally, our optimization target is to minimize the loss function $L = L_\text{margin} + \lambda L_\text{reg}$, where $\lambda$ is a hyper-parameter to balance these two terms.

To summarize, the HRCF is designed (1) to  avoid the overall embedding too close to the hyperbolic origin; (2) to take full advantage of hyperbolic space and; (3) to keep the internal dependencies. To be more specific, we first make a root alignment so that the root is placed in the correct hyperbolic position. Benefit from the self-origination, other nodes can be arranged in order, so the inherent structure can be kept; and then we apply a hyperbolic regularizer into the loss function, pushing nodes far away from the hyperbolic origin and taking full advantage of the hyperbolic space. 

\subsection{Theoretical Analysis}
In this section, we provide a theoretical analysis of two aspects. 
(A1) The embeddings with the tangential operations and the following exponential map operation will always be guaranteed to live in the tangent space at origin $\mathcal{T}_\mathbf{o}\mathbb{H}^n$ and hyperbolic space $\mathbb{H}^n$, respectively. 
(A2) The proposed HRCF is able to overcome the oversmoothing problem caused by the graph aggregation.

(A1) The initial embeddings $\mathbf{x}^{H,0}$ of all nodes are obtained by incorporating 0 with a sampled $\mathbf{x}^E$, then they can be interpreted as the points on the tangent space at origin $\mathcal{T}_\mathbf{o}\mathbb{H}^n$ based on {\sc Proposition}~\ref{prop:points with zero element}. After that, with the exponential map $\exp_\mathbf{o}(\cdot)$, the mapped nodes will just be placed in the hyperbolic space $\mathbb{H}^n$ without deviation as derived in~{\sc Proposition}~\ref{prop:exp_map_to_hyperbolic}. Next, using the logarithmic map, the embeddings are safely projected on the $\mathcal{T}_\mathbf{o}\mathbb{H}^n$ as the first element is 0. Then, the operations, including node aggregation (Equation~(\ref{equ:neighbors'aggregation_u},\ref{equ:neighbors'aggregation_i})), sum-pooling (Equation~(\ref{equ:multiple aggregation})), root computation~(Equation~(\ref{equ: embedding root})), and root alignment (Equation~(\ref{equ:alignment})) can always keep the first element as 0 in that these operations can be regarded as weighted sum. Thus the node always live in $\mathcal{T}_\mathbf{o}\mathbb{H}^n$ with Proposition~\ref{prop:lorentz_constraint}. Finally, the embeddings with $\exp_\mathbf{o}(\cdot)$ can be mapped back to $\mathbb{H}^n$ again according to Proposition~\ref{prop:exp_map_to_hyperbolic}.

\begin{prop}
\label{prop:points with zero element}
For any point $\{\mathbf{x}^E \in\mathbb{R}^{n+1} | \mathbf{x}_{[0]}^E=0, \mathbf{x}_{[1:n+1]}^E \in \mathbb{R}^{n}\}$. Then, $\mathbf{x}^E$ lives in the tangent space at origin $\mathcal{T}_\mathbf{o}\mathbb{H}^n, i.e, \mathbf{x}^E\in \mathcal{T}_\mathbf{o}\mathbb{H}^n$.
\end{prop}

\begin{prop}
\label{prop:lorentz_constraint}
For any point lives in $\mathbf{x}^\mathcal{T}\in\mathcal{T}_\mathbf{o}\mathbb{H}^n$, then $\mathbf{x}_{[0]}^\mathcal{T}=0.$
\end{prop}

\begin{prop}
\label{prop:exp_map_to_hyperbolic}
For any point $\mathbf{x}^\mathcal{T}\in \mathcal{T}_\mathbf{o}\mathbb{H}^n$, by applying the exponential map with reference point $\mathbf{o}$, i.e., $\exp_\mathbf{o}^K(\cdot)$, the mapped point $\mathbf{x}^H =\exp_\mathbf{o}^K(\mathbf{x}^\mathcal{T})=\exp_\mathbf{o}^K((0, \mathbf{x}^E))$ lives in~$\mathbb{H}_n^K$, or equivalently, is satisfied the Lorentz constraints, that is:
\begin{equation}
    \langle \mathbf{x}^H, \mathbf{x}^H \rangle_\mathcal{L} = -(x^H_0)^2+(x^H_1)^2 + \cdots + (x^H_n)^2 = -K
\end{equation}
\end{prop}

(A2) The proposed HRCF is capable of overcoming the issue of oversmoothing. In the following, we analyze the oversmoothing problem at first and then show why the proposed method can solve it. For simplicity, we especially focus on the neighborhood aggregation part in Equation~(\ref{equ:neighbors'aggregation_u},\ref{equ:neighbors'aggregation_i}).
\begin{theorem} [Shrinking Property in Hyperbolic Aggregation]
Let $\mathbf{x}_i^{\mathcal{T},\ell}\in\mathcal{T}_\mathbf{o}\mathbb{H}^{n}$, $\tilde{a}_{ij}=a_{ij}/d_{ii}$ be the normalized weight of edge $(i,j)$, and $D(X) = \sum_{\mathbf{x}_i,\mathbf{x}_j}\tilde{a}_{ij}\|\mathbf{x}_i^{\mathcal{T},{\ell}}-\mathbf{x}_j^{\mathcal{T},{\ell}}\|_2^2$ be a distance metric over all nodes and $\mathbf{x}_i^{\mathcal{T},{\ell+1}}=\sum_{j\in\mathcal{N}_{i}}\tilde{a}_{ij}\mathbf{x}_j^{\mathcal{T}, {\ell}}$. 
Then we have $D(X^{\mathcal{T}, \ell + 1})\leq D(X^{\mathcal{T},\ell}).$
\label{theorem: hyperbolic shrinking property}
\end{theorem}

Theorem~\ref{theorem: hyperbolic shrinking property} indicates that the distance of the connected nodes will shrink after the aggregation step in Equation~(\ref{equ:neighbors'aggregation_u},\ref{equ:neighbors'aggregation_i}). When we perform many aggregation operations, the connected nodes in the graph will shrink to a similar representation. At the same time, the nodes that are not directly connected (in the same connected component) will also shrink together due to their common connected nodes. Then the oversmoothing phenomenon will appear.

\begin{table}[!tp]
\caption{Statistics of the experimental datasets.}
\resizebox{0.40\textwidth}{!}{
\begin{tabular}{@{}lcccc@{}}
\toprule
\textbf{Dataset}      & \textbf{\#User} & \textbf{\#Item} & \textbf{\#Interactions} & \textbf{Density} \\ \midrule
\textbf{Amazon-CD}   & 22,947          & 18,395          & 422,301                 & 0.00100           \\
\textbf{Amazon-Book} & 52,406          & 41,264          & 1,861,118               & 0.00086          \\
\textbf{Yelp2020}     & 71,135          & 45,063          & 1,940,014               & 0.00047          \\ \bottomrule
\end{tabular}%
}
\label{tab:datasets}
\end{table}

\begin{table*}[!tp]
\caption{Comparison with various competing models. The best results are bold and the second best are underlined. Asterisks indicate that the Wilcoxon signed rank test for the difference between the best and second best models is statistically significant.}
\resizebox{0.98\textwidth}{!}{%
\begin{tabular}{@{}c|l|cc|cccc|cc|ccc|c@{}}
\toprule
\multicolumn{2}{l|}{Datasets}       & WRMF   & VAE-CF & TransCF   & CML    & LRML   & SML    & NGCF   & LightGCN & HAE    & HAVE   & HGCF   & HRCF(ours)   \\ \midrule 
\multirow{2}{*}{Amazon-CD}   & R@10 & 0.0863 & 0.0786 & 0.0518   & 0.0864 & 0.0502 & 0.0475 & 0.0758 & 0.0929   & 0.0666 & 0.0781 & \underline{0.0962} & \textbf{0.1003*} \\
                             & R@20 & 0.1313 & 0.1155 & 0.0791   & 0.1341 & 0.0771 & 0.0734 & 0.1150 & 0.1404   & 0.0963 & 0.1147 & \underline{0.1455} & \textbf{0.1503*} \\ \midrule
\multirow{2}{*}{Amazon-Book} & R@10 & 0.0623 & 0.0740 & 0.0407   & 0.0665 & 0.0522 & 0.0479 & 0.0658 & 0.0799   & 0.0634 & 0.0774 & \underline{0.0867} & \textbf{0.0900*} \\
                             & R@20 & 0.0919 & 0.1066 & 0.0632   & 0.1023 & 0.0834 & 0.0768 & 0.1050 & 0.1248   & 0.0912 & 0.1125 & \underline{0.1318} & \textbf{0.1364*} \\ \midrule
\multirow{2}{*}{Yelp2020}    & R@10 & 0.0470 & 0.0429 & 0.0247   & 0.0363 & 0.0326 & 0.0319 & 0.0458 & 0.0522   & 0.0360 & 0.0421 & \underline{0.0527} & \textbf{0.0537*}       \\
                             & R@20 & 0.0793 & 0.0706 & 0.0424   & 0.0638 & 0.0562 & 0.0544 & 0.0764 & 0.0866   & 0.0588 & 0.0691 & \underline{0.0884} & \textbf{0.0898*}  \\ \bottomrule
\end{tabular}%
}

\vspace{10pt}
\resizebox{0.98\textwidth}{!}{%
\begin{tabular}{@{}c|c|cc|cccc|cc|ccc|c@{}}
\toprule
\multicolumn{2}{c|}{Datasets}       & WRMF   & VAE-CF & TransCF  & CML    & LRML   & SML    & NGCF   & LightGCN & HAE    & HAVE   & HGCF   & HRCF(ours)      \\ \midrule
\multirow{2}{*}{Amazon-CD}   & N@10 & 0.0651 & 0.0615 & 0.0396   & 0.0639 & 0.0405 & 0.0361 & 0.0591 & 0.0726   & 0.0565 & 0.0629 & \underline{0.0751} & \textbf{0.0785*}    \\
                             & N@20 & 0.0817 & 0.0752 & 0.0488   & 0.0813 & 0.0492 & 0.0456 & 0.0718 & 0.0881   & 0.0657 & 0.0749 & \underline{0.0909} & \textbf{0.0947*}    \\ \midrule
\multirow{2}{*}{Amazon-Book} & N@10 & 0.0563 & 0.0716 & 0.0392   & 0.0624 & 0.0515 & 0.0422 & 0.0655 & 0.0780   & 0.0709 & 0.0778 & \underline{0.0869} & \textbf{0.0902*}    \\
                             & N@20 & 0.0730  & 0.0878 & 0.0474   & 0.0808 & 0.0626 & 0.055  & 0.0791 & 0.0938   & 0.0789 & 0.0901 & \underline{0.1022} & \textbf{0.1060*}    \\ \midrule
\multirow{2}{*}{Yelp2020}    & N@10 & 0.0372 & 0.0353 & 0.0214   & 0.0310 & 0.0287 & 0.0255 & 0.0405 & \underline{0.0461}   & 0.0331 & 0.0371 & 0.0458 & \textbf{0.0468*} \\
                             & N@20 & 0.0506 & 0.0469 & 0.0277   & 0.0428 & 0.0369 & 0.0347 & 0.0513 & 0.0582   & 0.0409 & 0.0465 & \underline{0.0585} & \textbf{0.0594*} \\ \bottomrule
\end{tabular}%
}
\label{table:results_Recall}
\end{table*}

In this work, we introduce a new term in the loss function. Then, it can be derived that when we minimize $L_\mathrm{reg}$ in Equation~(\ref{equ:x_norm}), or equivalently maximize 
the overall embedding norm $\bar{x}_\mathrm{norm}$, then the distance between node pair will be enlarged to some extent, which prevents the distance decay caused by the graph aggregation in the tangent space $\mathcal{T}_\mathbf{o}\mathbb{H}^n$. 
Thus, the proposed method is able to overcome the oversmoothing problem in the theoretical aspect, which is formally presented in Proposition~\ref{prop:overcome_shrinking}. In section~(\ref{sec:layer_analysis}), we conduct a detailed experimental analysis to verify the theoretical results.

\begin{prop}
\label{prop:overcome_shrinking}
By minimizing the proposed loss $L_{\mathrm{reg}}$ in Equation~(\ref{equ:x_norm}), or equivalently maximize the average embedding norm $\bar{x}_\mathrm{norm}$, then the distance $\sum_{i,j\in V}\|\mathbf{x}_i-\mathbf{x}_j\|_2^2$ over all node pairs will be enlarged.
\end{prop}

\subsection{Geometric Analysis}
Furthermore, we show the benefits of pushing the nodes far away from the origin from the geometric aspect. The similar conclusion has been appeared in previous works~\cite{HNN,HyperML2020} on Poincar\'e Ball model. In hyperboloid manifold, the analysis is the same. Let us consider three points in hyperbolic space, which are two points $\mathbf{x}^H$ and $\mathbf{y}^H$ with $\|\mathbf{x}^H\|=\|\mathbf{y}^H\|=a(a>0)$ and the origin $\mathbf{o}$. Assuming the origin $\mathbf{o}$ is the parent of $\mathbf{x}^H$ and $\mathbf{y}^H$. As the points $\mathbf{x}^H$ and $\mathbf{y}^H$ parallel move towards the area far away from the origin, the ratio between the hyperbolic distance $d_\mathcal{L}(\mathbf{x}^H, \mathbf{y}^H)$ and the sum of the HDO of $\mathbf{x}^H$ and $\mathbf{y}^H$, i.e., $\frac{d_\mathcal{L}(\mathbf{x}^H, \mathbf{y}^H)}{d_\mathcal{L}(\mathbf{x}^H, \mathbf{o}) + d_\mathcal{L}(\mathbf{y}^H, \mathbf{o})}$ approaches 1, or equivalent, $d_\mathcal{L}(\mathbf{x}^H,\mathbf{y}^H)$ approaches $d_\mathcal{L}(\mathbf{x}^H, \mathbf{o}) + d_\mathcal{L}(\mathbf{y}^H, \mathbf{o})$. However, in Euclidean space, the ratio is always the same since the Euclidean space is a flat space and the movement cannot change the ratio.
Overall, pushing nodes to approach the boundary can facilitate preserving the real hyperbolic distance and reducing the distortion.

\section{Experiments}

\subsection{Experimental Settings}
\textbf{Datasets.} In this work, we use three publicly available datasets Amazon-Book$^1$, Amazon-CD\footnote{https://jmcauley.ucsd.edu/data/amazon/}, and Yelp2020\footnote{https://www.yelp.com/dataset}, which are also employed in the HGCF~\cite{sun2021hgcf}. The statistics are summarized in  \tableautorefname~\ref{tab:datasets}.\footnote{Note that we use the same datasets and preprocessing as HGCF. There is a slight difference in the number of users in Yelp2020, which is a typo in HGCF, please refer to the issue https://github.com/layer6ai-labs/HGCF/issues/2} {Amazon-CD} and {Amazon-Book} are two widely-used product recommendation datasets released by Amazon. {Yelp2020} is the 2020 edition of the
Yelp challenge, where the local businesses such as restaurants and
bars are regarded as items.
In these datasets, the ratings are converted into binary preferences by applying a threshold $\geq4$ which simulates the implicit feedback setting~\cite{ma2020probabilistic,sun2021hgcf}. 
For each dataset, we randomly split historical interactions into $80/20\%$ for training and test.

\textbf{Compared methods.} 
To verify the effectiveness of our proposed method, the compared  methods include both well-known or leading hyperbolic models and Euclidean baselines. For hyperbolic models, the HGCF~\cite{sun2021hgcf}, HVAE and HAE are compared. Besides, we include some recent strong Euclidean baselines, such as LightGCN~\cite{he2020lightgcn} and NGCF~\cite{wang2019ngcf}. Last but not least, factorization-based models, WRMF~\cite{wrmf2008} and VAE-CF~\cite{VAECF2018}; and metric learning-based models, TransCF~\cite{park2018collaborative}, CML~\cite{CML2017}, LRML~\cite{LRML2018}, and SML~\cite{SML2020}, are also considered. The data preprocessing and running environments of Amazon-Books, Amazon-CDs, and Yelp2020 are the same as the HGCF, so we refer to the results of the baseline from the HGCF.\footnote{Note that the R@10 of HGCF on Yelp is 0.0527 with official code (\textit{c.f.} footnote 7).}

\textbf{Experimental setup.}
To reduce the experiment workload and keep the comparison fair, we closely follow the settings of HGCF.\footnote{https://github.com/layer6ai-labs/HGCF.}
More specifically, the embedding size is set as 50 and the total training epochs are fixed as 500. The $\lambda$ in the loss function is in the range of $\{10, 15, 20, 25, 30\}$ and the aggregation order is searched from 2 to 10. For the margin, we search it in the scope of $\{0.1, 0.15, 0.2\}$. We utilize the Riemannian SGD~\cite{bonnabel2013stochastic} with weight decay in the range of $\{1e-4, 5e-4,1e-3\}$ to learn the network parameters where the learning rates are in $\{0.001, 0.0015, 0.002\}$. RSGD mimics the stochastic gradient descent optimization while taking into account the geometry of the hyperbolic manifold~\cite{bonnabel2013stochastic}. For the baseline settings, we refer to \cite{sun2021hgcf}.

\textbf{Evaluation metrics}.
To evaluate the effectiveness of top-K recommendation and preference ranking, we adopt two widely-used evaluation protocols: Recall and NDCG ~\cite{ying2018graph}. For each observed user-item interaction, we treat it as a positive instance, and then conduct the negative sampling strategy to pair
it with one negative item that the user did not rate before.

\textbf{Model and time complexity}. 
As for the time complexity, we analyze it from the training and test phase. In the training phase, we neglect the $\ell$-order multiplication and addition since it can be computed before the training phase and then the training time complexity is $N^2d + 3N + 2|E_{train}|$ for our method, and $N^2d + 2N + 2|E_{train}|$ for HGCF, where $N^2d$ is the time complexity of matrix multiplication, $2N$ in HGCF is originated from the exponential and logarithmic map, one more $N$ in our method is from the hyperbolic regularization, and $2|E_{{train}}|$ is for the hyperbolic margin ranking loss on training positive and negative interactions where $|E_{{train}}|$ ($|E_{{test}}|$) is the number of training (test) interactions. In the test phase, the time complexity of our method is the same as HGCF that is $|E_{test}|$. It is worth mentioning that the time complexity can be further reduced via parallel computing on GPU. In summary, the introduction of hyperbolic geometric regularization brings negligible computation costs. 

\subsection{Overall Performance}
The empirical results are reported in Table
\ref{table:results_Recall}, where the best results are in bold and the second best are underlined. In summary, HRCF comprehensively outperforms the baseline models on all datasets across both Recall@K and NDCG@K, demonstrating the effectiveness of our proposal. We further have the following observations. First, hyperbolic models equipped with ranking loss (i.e., HGCF and HRCF) show significant advantage compared with their Euclidean counterparts (i.e., LightGCN), demonstrating the superiority of hyperbolic geometry for modeling user-item networks. It is also noted that the cross-entropy loss powered two hyperbolic baselines (i.e., HAE and HVAE) are barely satisfactory, which shows the importance of the task-specific loss function. Last but not least, our proposal confirms the power of graph neural network based collaborative filtering as it further enlarges the gaps of that with the MF-based and metric-based methods. 

\begin{figure*}[h]
\centering
\includegraphics[width=4.251cm]{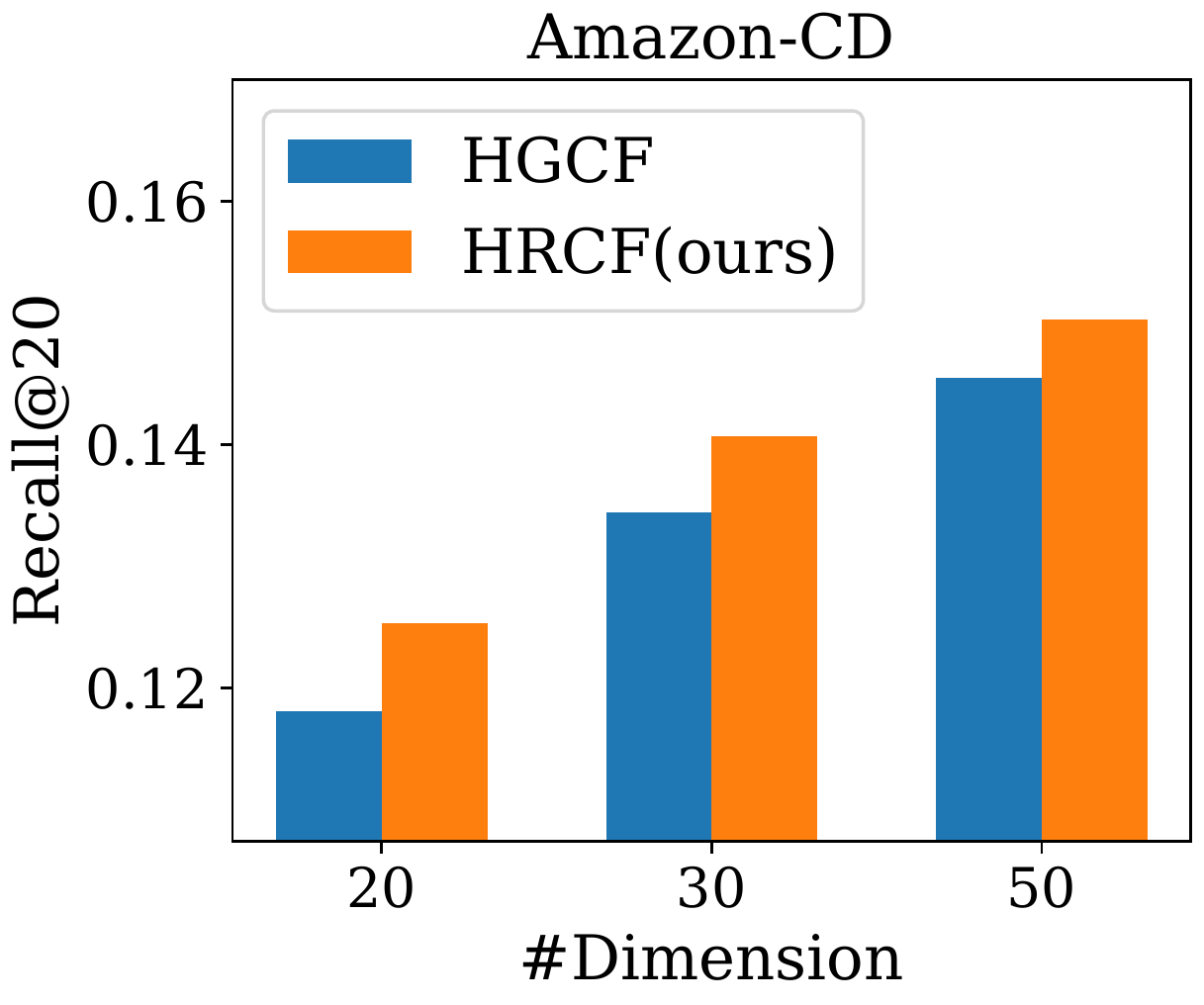}
\includegraphics[width=4.251cm]{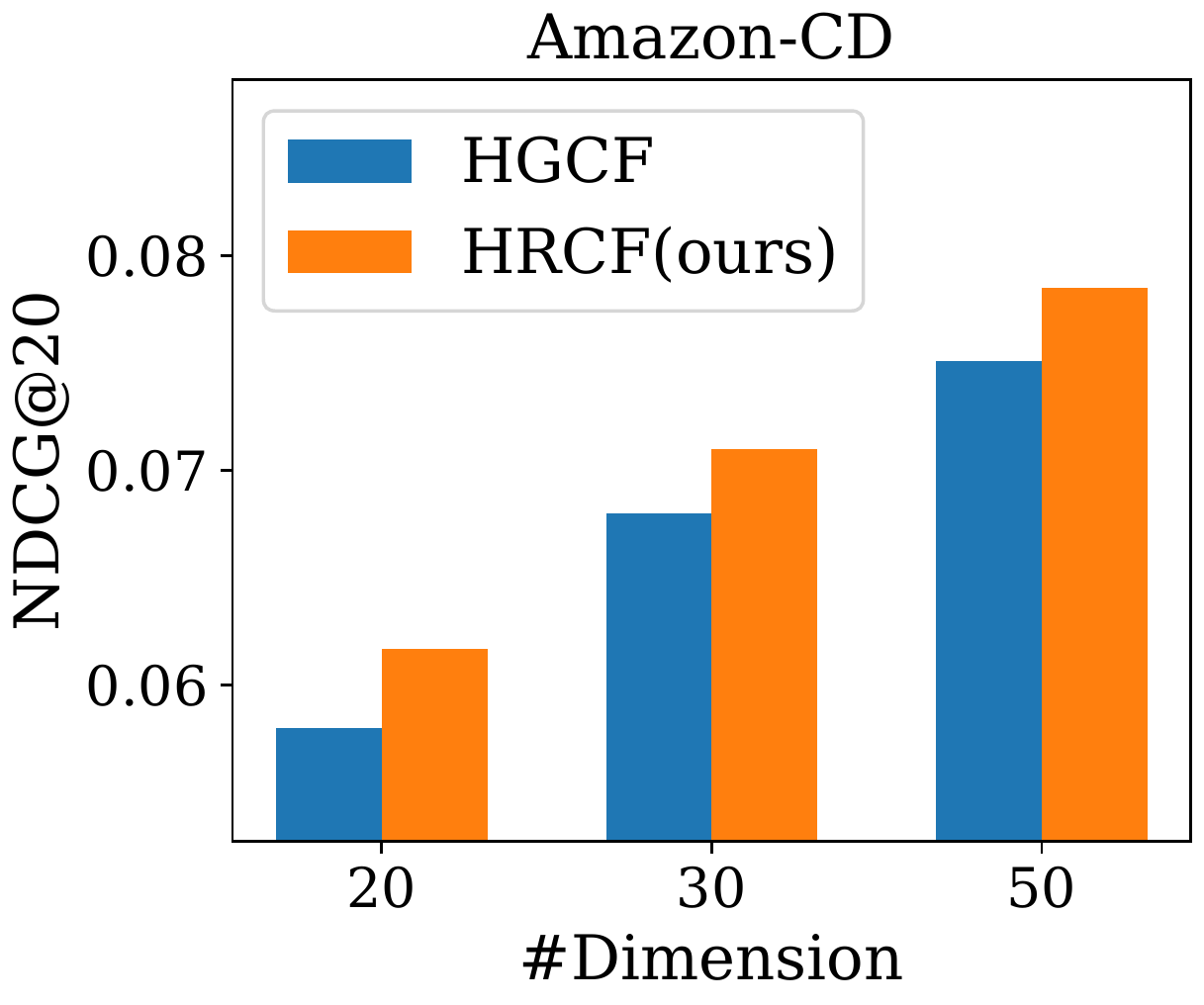}
\includegraphics[width=4.251cm]{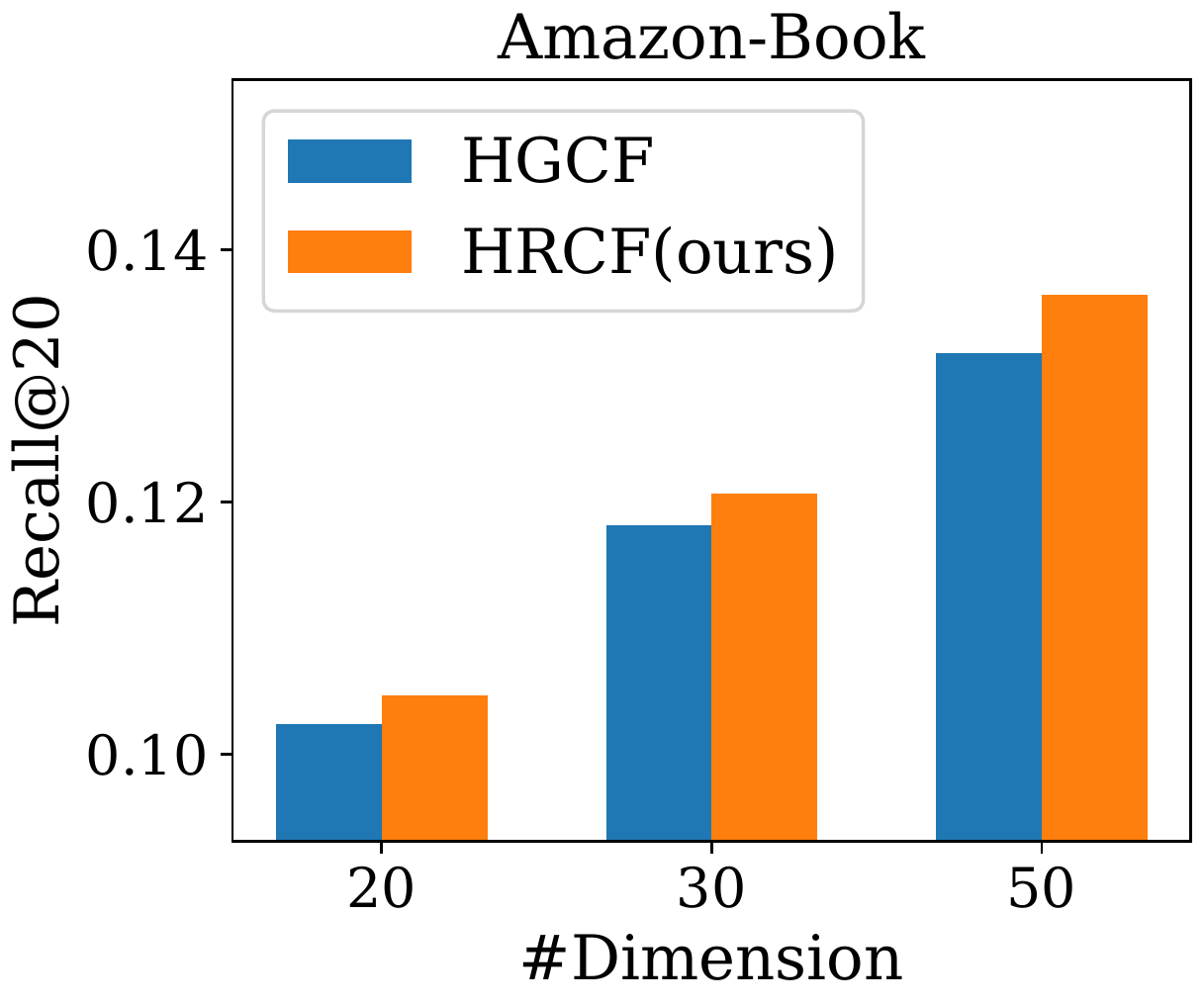}
\includegraphics[width=4.251cm]{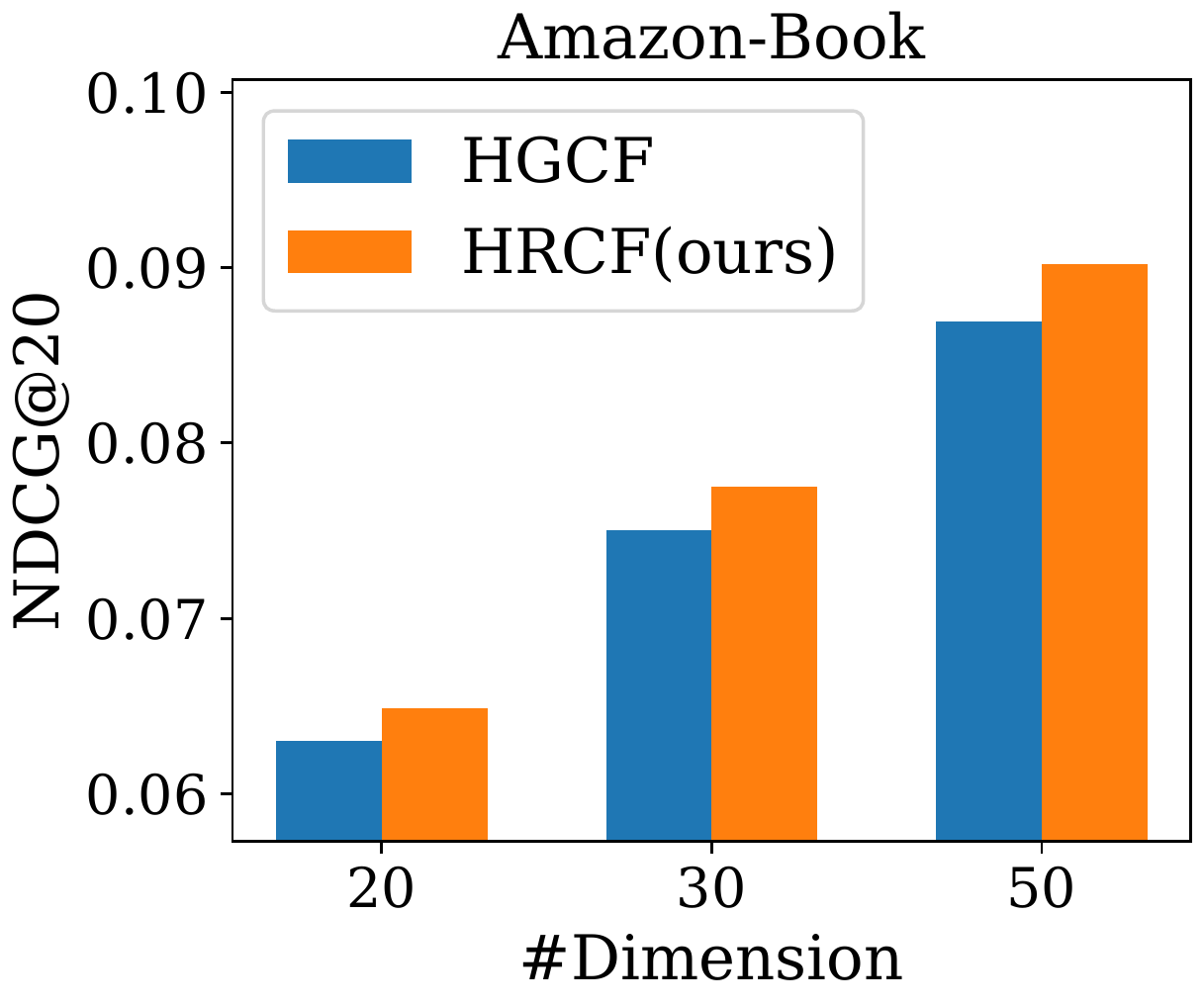}
\caption{Performance with different embedding dimensions on Amazon-CD and Amazon-Book datasets. The comparisons are with dimensions 20,30 and 50. The evaluation metrics are Recall@20, NDCG@20, and other metrics present similar results.}
\label{fig:performance with different dimension}
\vspace{-10pt}
\end{figure*}
\begin{figure*}[!t]
\centering
\includegraphics[width=4.20cm]{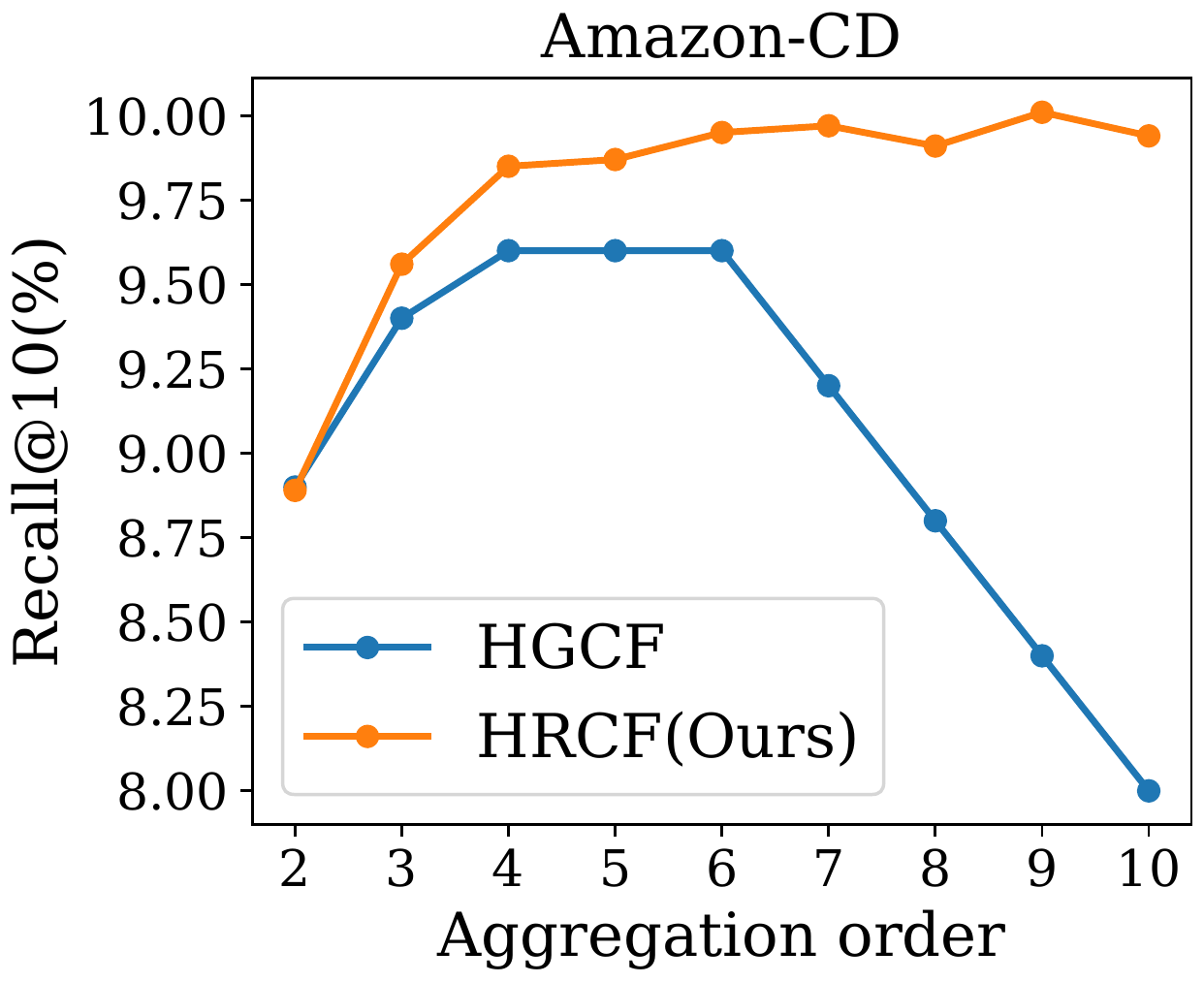}
\includegraphics[width=4.20cm]{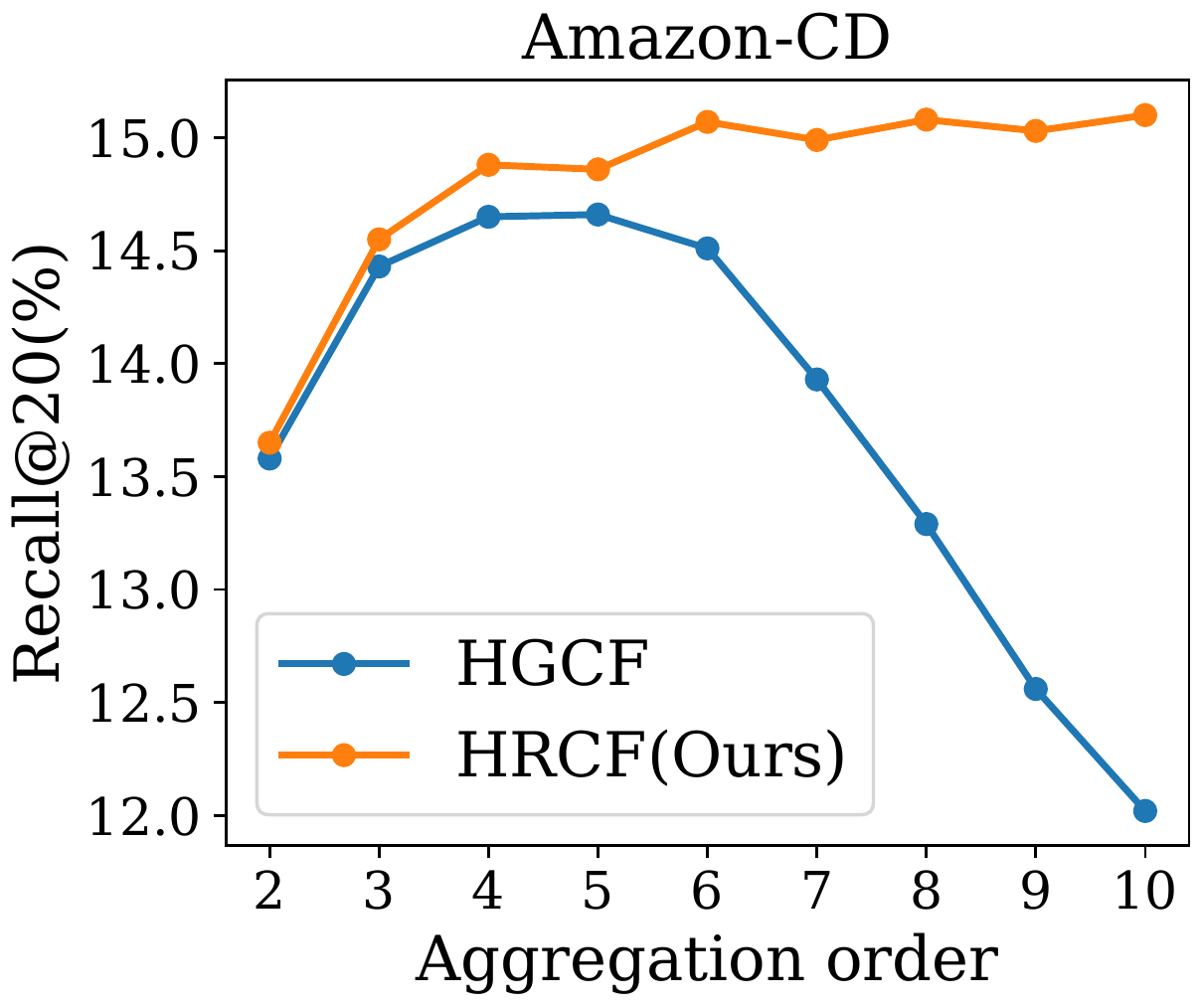}
\includegraphics[width=4.20cm]{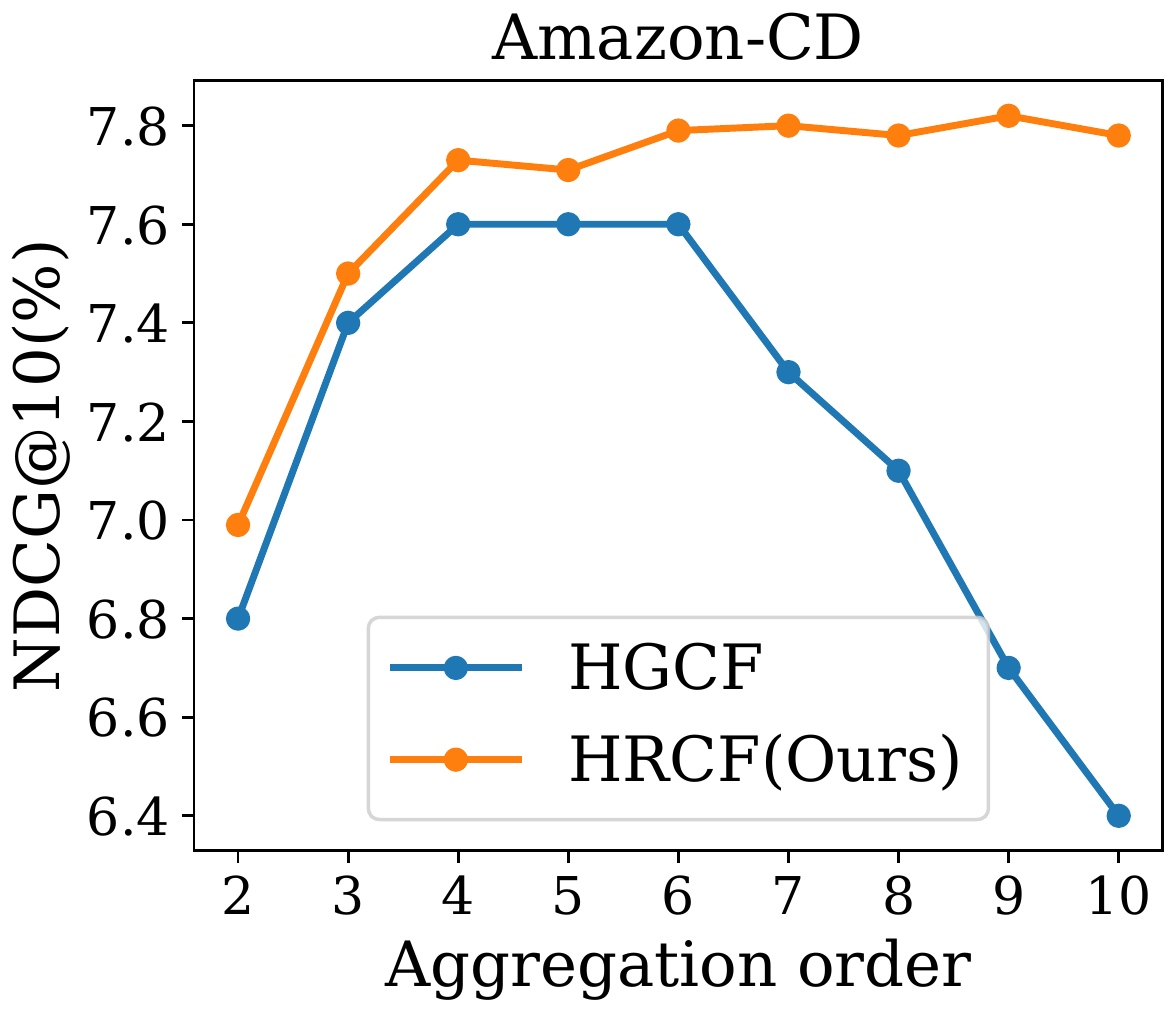}
\includegraphics[width=4.20cm]{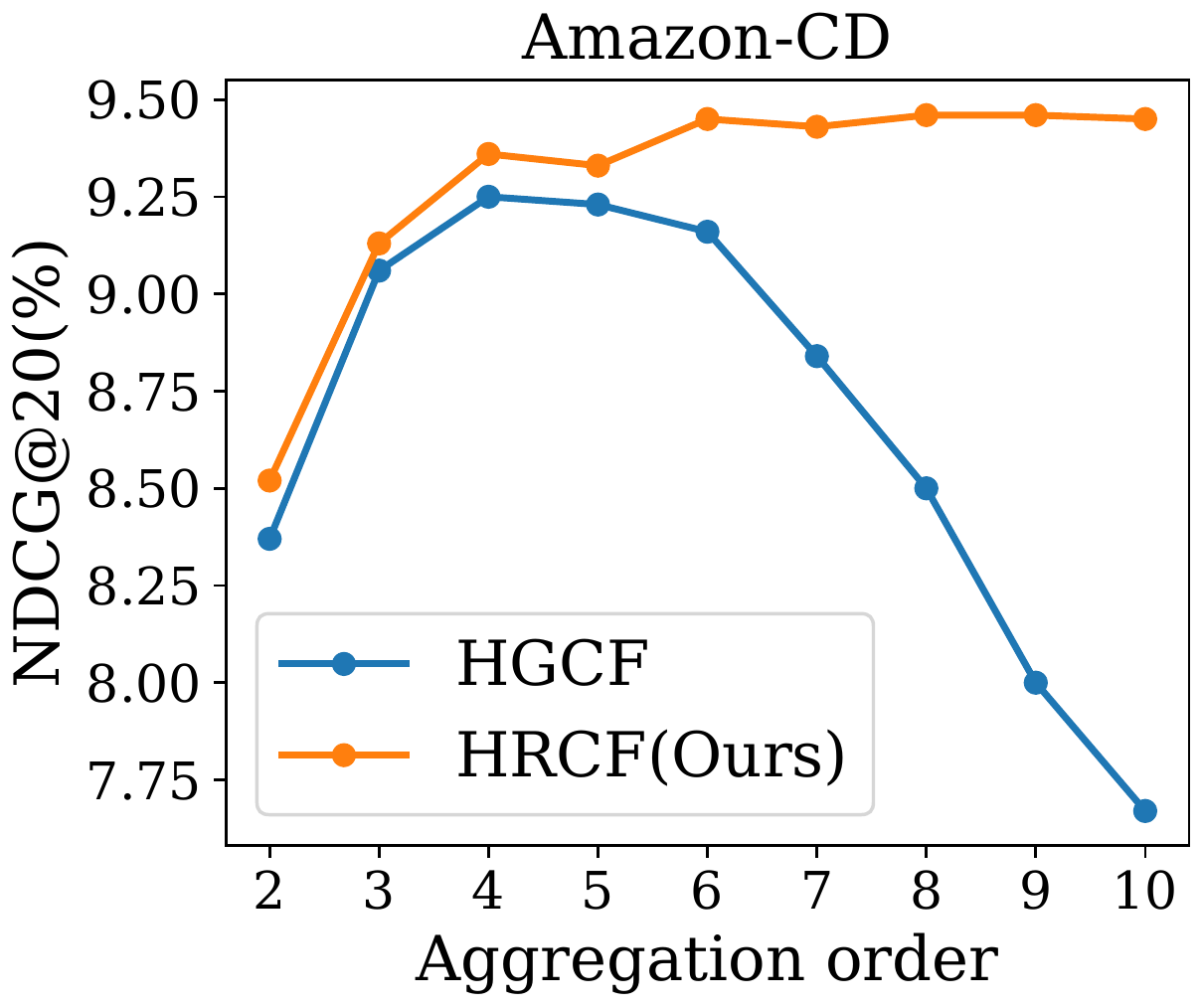}

\includegraphics[width=4.20cm]{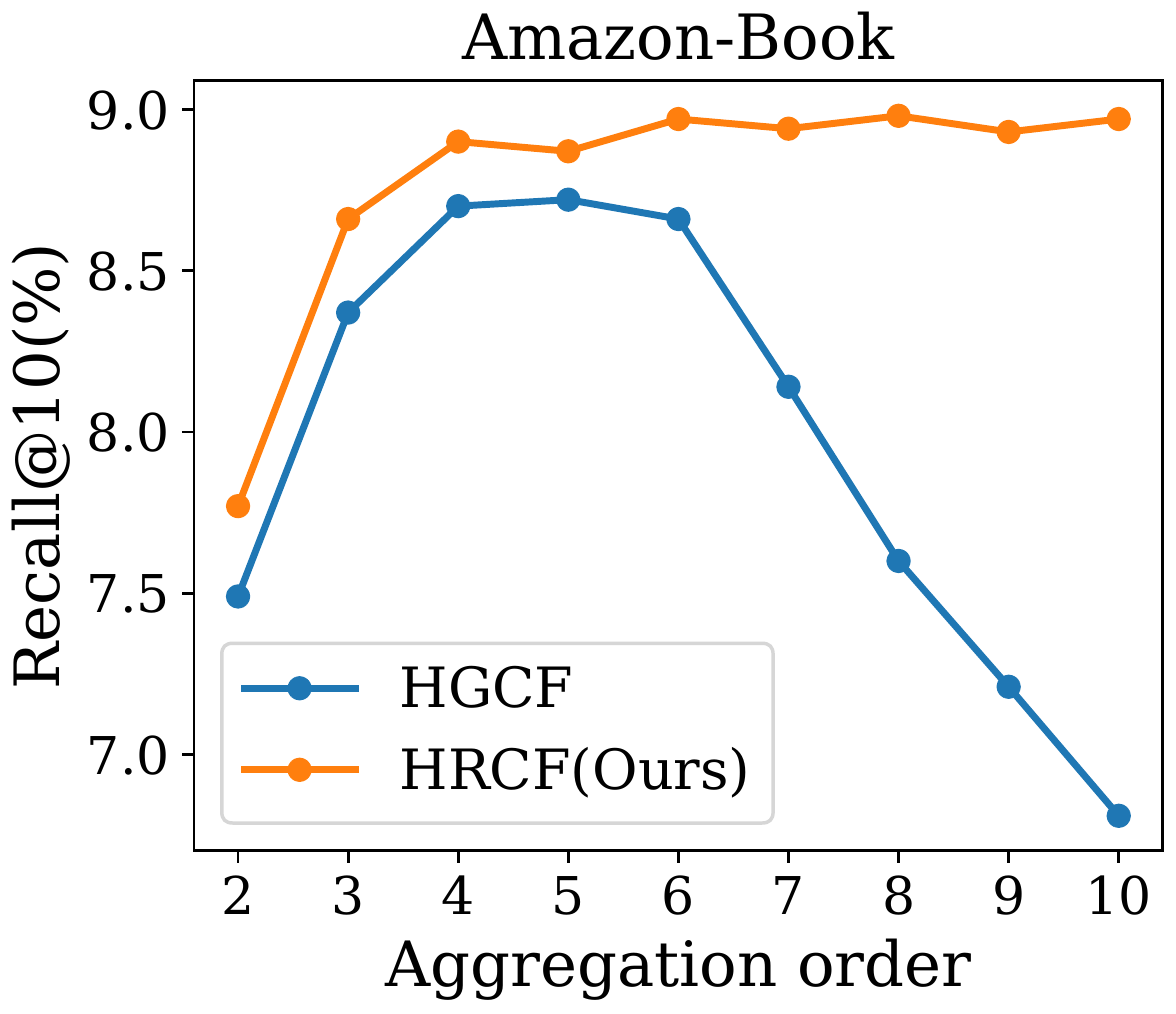}
\includegraphics[width=4.20cm]{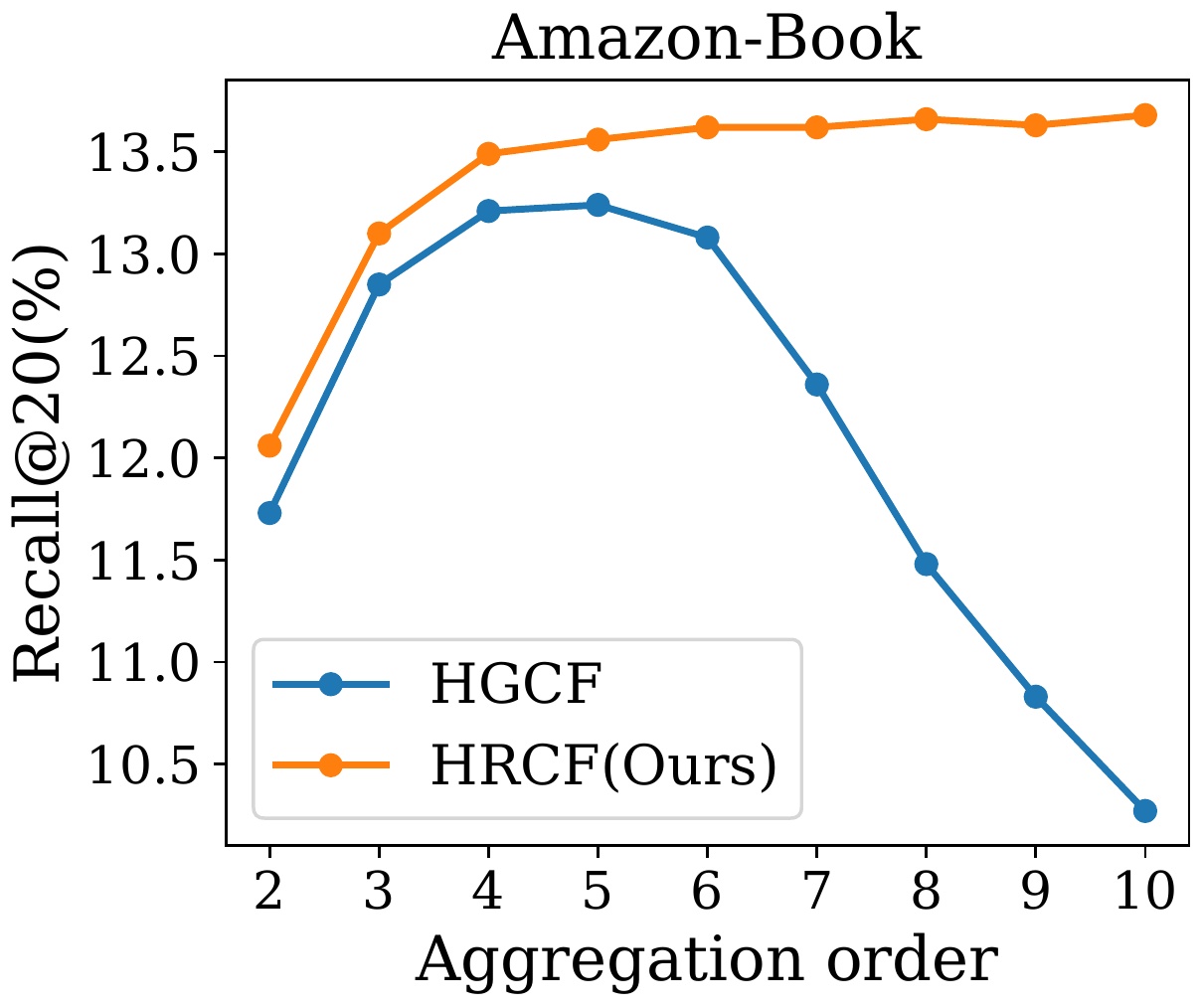}
\includegraphics[width=4.20cm]{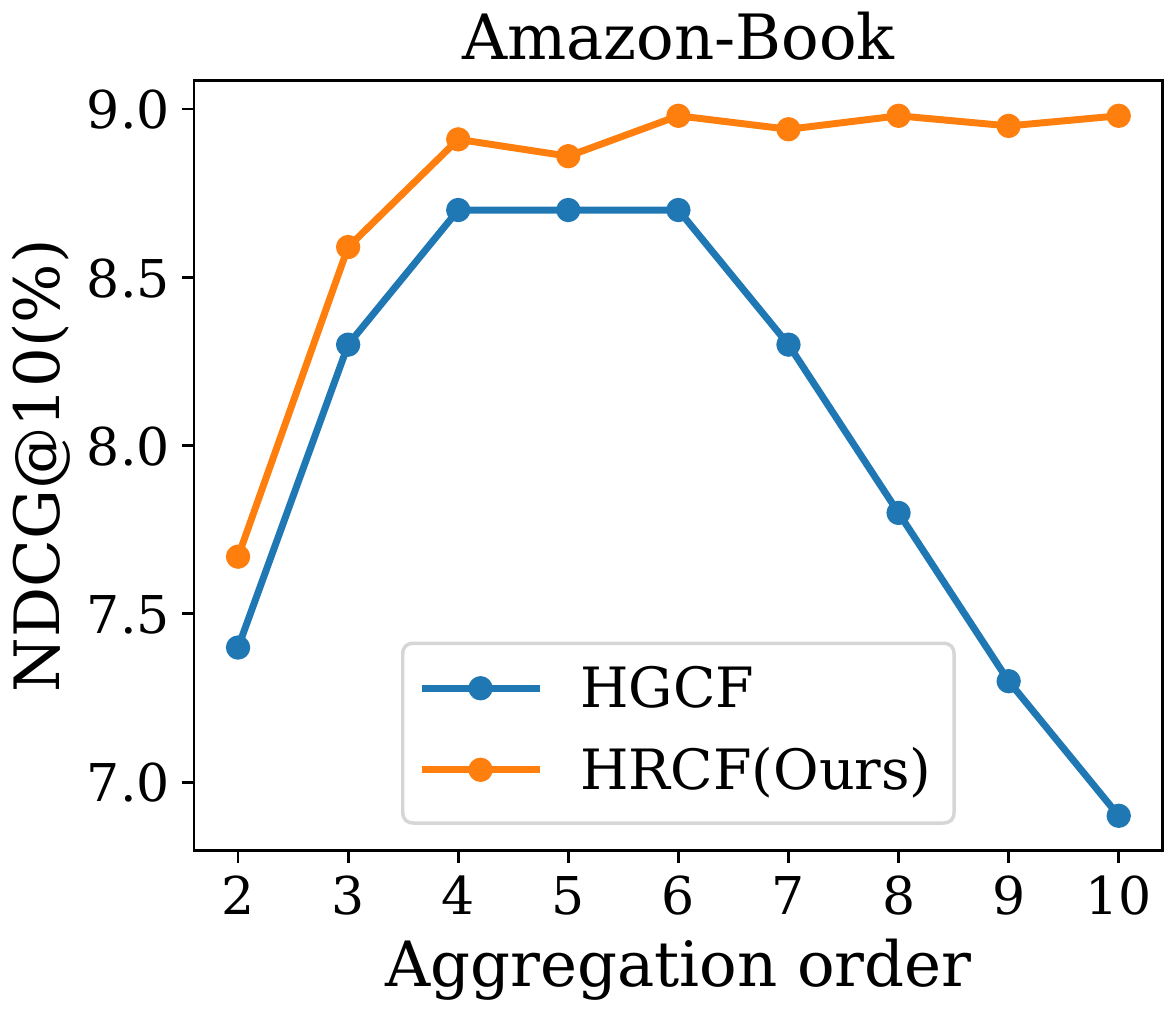}
\includegraphics[width=4.20cm]{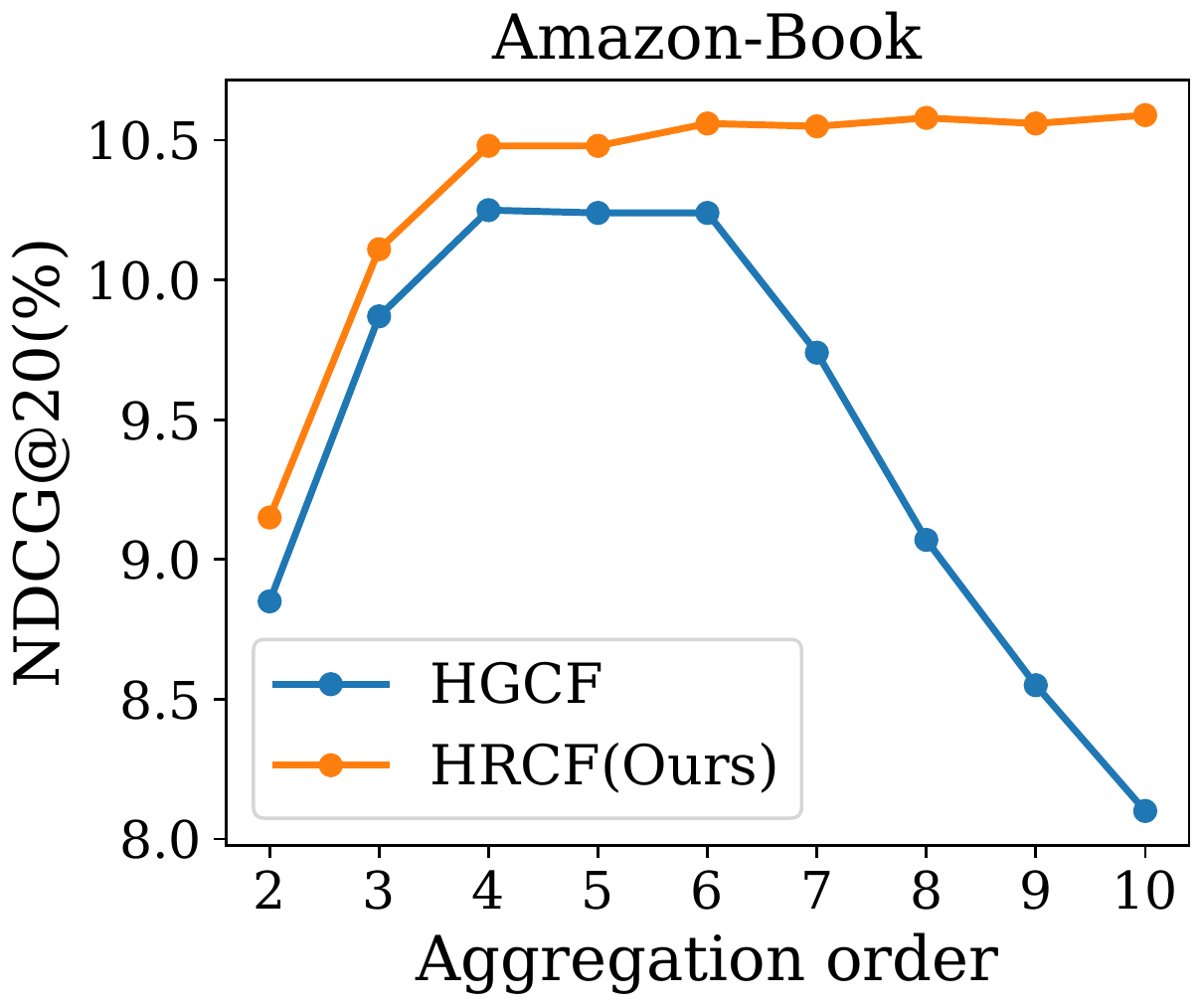}
\caption{Performance with different aggregation orders on Amazon-CD and Amazon-Book datasets. The comparisons range from order 2 to 10 and the evaluation metrics are Recall@10, Recall@20, NDCG@10,and NDCG@20.}
\label{fig:order-comparsion}
\vspace{-10pt}
\end{figure*}
\subsection{Lightweightness w.r.t. Embedding Size}
Table \ref{table:results_Recall} demonstrates the superiority of our proposal with the embedding dimension 50. 
We here further evaluate the model performance with lower embedding dimensions size, which is particularly prominent if the computation and storage resources are limited. Specifically, we reduce the dimension size from 50 to 20 and record the performance of our proposed HRCF and the second-best HGCF, which are further displayed in Figure~\ref{fig:performance with different dimension}.  We observe that the results are highly correlated with the embedding size and the smaller embedding size reduces the computation burden but degrades the performance as well. However, HRCF continuously outperforms the HGCF for all the scenarios and the advantage is even more obvious in the case of Amazon-CD with the smallest dimension size 20. 
This further confirms the HRCF’s advantage for the lightweight recommender system where the model complexity is limited by computing and storage resources.

\begin{table}[!tp]
\caption{Recall@10 and NDCG@10 on Amazon-CD/Book datasets as
training epoch varies from 100 to 20.}
\resizebox{0.46\textwidth}{!}{%
\begin{tabular}{@{}lccccccc@{}}
\toprule
\multirow{2}{*}{Model}       & Datasets  & \multicolumn{3}{c}{Amazon-CD} & \multicolumn{3}{c}{Amazon-Book} \\ \cmidrule(l){2-8} 
                             & Epoch     & 100      & 60       & 20      & 100       & 60       & 20       \\ \midrule
\multirow{2}{*}{HGCF}        & Recall@10 & 0.0880   & 0.0820   & 0.0680  & 0.0712    & 0.0644   & 0.0505   \\
                             & NDCG@10   & 0.0690   & 0.0640   & 0.0530  & 0.0700    & 0.0630   & 0.0490   \\ \midrule
\multirow{2}{*}{HRCF}        & Recall@10 & 0.0928   & 0.0899   & 0.0792  & 0.0786    & 0.0727   & 0.0589   \\
                             & NDCG@10   & 0.0716   & 0.0692   & 0.0603  & 0.0780    & 0.0716   & 0.0567   \\ \midrule
\multirow{2}{*}{Rel.Improv.} & Recall@10 & 5.45\%     & 9.63\%     & 16.47\%   & 10.39\%     & 12.89\%    & 16.63\%    \\
                             & NDCG@10   & 3.77\%     & 8.12\%     & 13.77\%   & 11.43\%     & 13.65\%    & 15.71\%    \\ \bottomrule
\end{tabular}%
}
\label{tab:performance-epoch-study}
\vspace{-10pt}
\end{table}
\subsection{Robustness w.r.t Aggregation Order}
\label{sec:layer_analysis}
One of the main advantages of collaborative filtering using graph convolutional networks is its ability to extract high-order relationships, which is achieved by the multi-order aggregation in graph convolution as given by Equation~(\ref{equ:neighbors'aggregation_u},\ref{equ:neighbors'aggregation_i}). Due to the property of weight decay as demonstrated in Theorem ~\ref{theorem: hyperbolic shrinking property}, all nodes tend to be similar and smooth after the message aggregation. However, to extract high-order dependencies, it is inevitable to apply graph aggregation multiple times and the oversmoothing problem will appear. Based on our theoretical analysis, the proposed HRCF can well overcome this problem. Here we have carried out a series of comprehensive experiments to verify the theoretical results. Figure~\ref{fig:order-comparsion} illustrates the comparison between HGCF and the proposed HRCF with the metric Recall@10, Recall@20, NDCG@10 and NDCG@20. As illustrated in Figure~\ref{fig:order-comparsion}, the performance curve of HRCF on four different evaluation metrics consistently outperforms the baseline HGCF, which indicates the superiority of the proposed HRCF. 
Furthermore, it can be found that the performance of HRCF increases at first and then tends to be stable as the aggregation order increases; while for the baseline HGCF, it suddenly drops when the aggregation order reaches a threshold. On the order hand, the performance of HRCF increases or saturates with the aggregation order. It shows the ability of HRCF for deeper layers which is of importance for complex and large-scale datasets ~\cite{chen2020simple}. 

\subsection{Convergent Speed w.r.t Training Epoch}
Likewise, we study the efficacy of the proposed method using fewer epochs for training the proposed model HRCF and the most competitive counterpart HGCF. Table~\ref{tab:performance-epoch-study} lists the performance records with epochs 100, 60 and 20 on metrics Recall@10 and Recall@20. Based on the results, we have the following conclusions: (1) the proposed HRCF repeatedly outperforms the baseline models across different epochs; (2) with a few epochs, the proposed method obtains a larger relative improvement (\textit{c.f.} Rel.Improv.), which shows our proposed method can speed up the convergence of the training process. 

\section{Conclusion}
Hyperbolic space is a curved space with constant negative curvature, of which the volume increases exponentially. Therefore, it is much more spacious and roomy than the flat Euclidean space. In particular, it is proficient at matching such data to a hierarchical structure or pow-law distribution. To improve the utilization of hyperbolic space and enhance the performance of a hyperbolic-powered recommendation system, we propose a simple yet effective method, HRCF. The main idea of HRCF is to push the overall embeddings in the hyperbolic space far away from the origin to make the most of the roomier space while maintaining the latent structure. To achieve this goal, we propose to use the embedding midpoint as the root node and align it with the hyperbolic origin. Furthermore, we maximize the norm of overall embeddings in the loss function. By analyzing the property of our proposal, we theoretically proved that our method improves the model performance by enforcing the preferred or non-preferred items to be more discrepancy, which alleviates the oversmoothing of graph-based models and provides a promising solution for the deeper hyperbolic graph neural collaborative filtering. It is of great importance to the complex and large-scale user-item datasets.  The experimental results demonstrate the effectiveness of our proposal, as it achieves highly competitive results and outperforms both leading graph-based hyperbolic and Euclidean models. The empirical results also show our HRCF is a robust, deeper, and lightweight neural graph collaborative filtering model. Note that the proposed idea is decoupled from the collaborative filtering, and theoretically,
they can be applied to other hyperbolic models. In future work, we will extend our idea to more recommendation scenes or non-recommendation scenarios, like the hyperbolic temporal link prediction~\cite{yang2021discrete}.

\section*{Acknowledgements}
The work described in this paper was partially supported by the National Key Research and Development Program of China (No. 2018AAA0100204) and Research Grants Council of the Hong Kong Special Administrative Region, China (CUHK 2300174, Collaborative Research Fund, C5026-18GF. We would like to thank the anonymous reviewers for their constructive comments.
\newpage
\bibliographystyle{ACM-Reference-Format}
\bibliography{reference}
\newpage
\appendix

\section*{Appendix}
\section{Performance on Head and Tail Items}
To have a better understanding of the effectiveness of our method, we further check the performance of HRCF on the head and tail item. Specifically, we sort by the degree of the items in descending order. The top 20\% of the items are classified as head items which are popular and liked by numerous users, while the remaining are grouped as the tail items which are unpopular items indicating some personalized items or some new items. Although we are aware that this split strategy cannot precisely separate popular and unpopular items, it adequately depicts the findings to a certain degree.
The results of head and tail items are recorded in Table~\ref{tab:performance-of-head-and-tail}. 

The findings indicate that the proposed model HRCF can boost recommendation for both head and long-tail items, with a much more substantial impact on tail items than head items, which is consistent with our motivation and confirms the effectiveness of the proposal. 
The noticeable improvements on tail items are intuitively understandable. The main reason is that the proposed method attempts to optimize a number of long-tail nodes to the region that are far away from the origin and has more spacious room, enabling the tail nodes to separate from one another very well, thereby gaining obviously improved performance on tail items.
In comparison to the most popular items at the head, these tail items are more critical for personalized recommendations since they can better reflect the personalized interests of the user.

\begin{table}[h]
\caption{Performance of head and tail items with Recall@10 and NDCG@10 on Amazon-CD and Amazon-Book.}
\label{tab:performance-of-head-and-tail}
\resizebox{0.46\textwidth}{!}{%
\begin{tabular}{@{}l|cccc|cccc@{}}
\toprule
\multirow{2}{*}{Datasets} & \multicolumn{4}{c|}{Amazon-CD}                       & \multicolumn{4}{c}{Amazon-Book}                     \\ \cmidrule(l){2-9} 
                          & \multicolumn{2}{c}{Head} & \multicolumn{2}{c|}{Tail} & \multicolumn{2}{c}{Head} & \multicolumn{2}{c}{Tail} \\ \midrule
Metric & R@10        & N@10       & R@10        & N@10       & R@10        & N@10       & R@10        & N@10       \\ \midrule
HGCF   & 0.0667      & 0.0553     & 0.0295      & 0.0201     & 0.0550      & 0.0578     & 0.0317      & 0.0291     \\
Ours   & 0.0674      & 0.0558     & 0.0329      & 0.0227     & 0.0563      & 0.0593     & 0.0337      & 0.0309     \\
Gain(\%)   & +1.05        & +0.90       & +11.53       & +12.94      & +2.36        & +2.60       & +6.31        & +6.19       \\ \bottomrule
\end{tabular}%
}
\vspace{-10pt}
\end{table}

\section{Proof}
\subsection{Proof of Proposition~\ref{prop:points with zero element}}
\begin{proof}
Based on the Definition~\ref{def:lorentz_tangent_space}, the inner product of any point $\mathbf{x}^E$ at the hyperbolic origin point $\mathbf{o}=(1,\mathbf{0}_{[1:n+1]})$ is given by,
    \begin{equation}
        \begin{aligned}
        \langle \mathbf{x}^E,\mathbf{o} \rangle_\mathcal{L} = \langle(0, \mathbf{x}_{[1:n+1]}^E), (1,\mathbf{0}_{[1:n+1]})\rangle_\mathcal{L} = 0,
        \end{aligned}
    \end{equation}
and thus $\mathbf{x}^E$ lives in the tangent space at origin $\mathcal{T}_\mathbf{o}\mathbb{H}^n$, $i.e, \mathbf{x}^E\in \mathcal{T}_\mathbf{o}\mathbb{H}^n$. 
\end{proof}

\subsection{Proof of Proposition~\ref{prop:lorentz_constraint}}
\begin{proof}
For any point $\mathbf{x}^\mathcal{T}$ lives in the tangent space at origin, we have the following deviation:
\begin{equation}
    0=\langle\mathbf{x}^\mathcal{T},\mathbf{o} \rangle_\mathcal{L}=-\mathbf{x}_0^\mathcal{T}\cdot1 + \sum_i^n \mathbf{x}_i^\mathcal{T}\cdot 0=-\mathbf{x}_0^\mathcal{T}.
\end{equation}
Thus, we know that $\mathbf{x}_0^\mathcal{T}=0$.
\end{proof}
\subsection{Proof of Proposition~\ref{prop:exp_map_to_hyperbolic}}

\begin{proof}
According to the definition of Lorentz inner product, we have the following:
    \begin{equation}
        \begin{aligned}
            &\langle \mathbf{x}^H, \mathbf{x}^H \rangle_\mathcal{L}  = \langle \exp_\mathbf{o}^K((0, \mathbf{x}^E)), \exp_\mathbf{o}^K((0, \mathbf{x}^E)) \rangle_\mathcal{L} \\
            &=\|(\sqrt{K}\cosh(\alpha), \sqrt{K}\sinh(\alpha)\frac{\mathbf{x}^E}{\|\mathbf{x}^E\|_2} )\|_\mathcal{L}^2\\
            &=-(\sqrt{K}\cosh(\alpha))^2 +  (\sqrt{K}\sinh(\alpha))^2\frac{\|\mathbf{x}^E\|_2^2}{\|\mathbf{x}^E\|_2^2}\\
            & = -K\cosh^2(\alpha) + K\sinh^2(\alpha)\\
            & = -K(\cosh^2(\alpha)-\sinh^2(\alpha)) \\
            & = -K,
        \end{aligned}
    \end{equation}
where $\alpha = {\|\mathbf{x}^E\|_2}/{\sqrt{K}}$ and $K=1$ in this work. Then, we know that when applying the exponential map with reference point $\mathbf{o}$, i.e., $\exp_\mathbf{o}^K(\cdot)$, the mapped point $\mathbf{x}^H =\exp_\mathbf{o}^K(\mathbf{x}^\mathcal{T})=\exp_\mathbf{o}^K((0, \mathbf{x}^E))$ always lives in~$\mathbb{H}_n^K$, or equivalently, is satisfied the Lorentz constraints.
\end{proof}

\subsection{Proof of Theorem~\ref{theorem: hyperbolic shrinking property}}
\begin{proof}
The aggregation is applied on the tangent space at origin, according to {\sc Proposition~\ref{prop:lorentz_constraint}}, we know that the value of zero-th coordinate $\mathbf{x}_{i[0]}^\ell=0$, thus $D(X) = \sum_{\mathbf{x}_i, \mathbf{x}_j}\tilde{a}_{ij}\|\mathbf{x}_i^{\ell}-\mathbf{x}_j^{\ell}\|_2^2=D(X_{[1:n+1]})$. 
Based on the Theorem 4 given by ~\citeauthor{wang2020unifying}~\cite{wang2020unifying}, we know that $D(X_{[1:n+1]}^{\mathcal{T},\ell + 1})\leq D(X_{[1:n+1]}^{\mathcal{T},\ell})$. Therefore, $D(X^{\ell + 1})\leq D(X^{\mathcal{T},\ell})$ holds on.
\end{proof}

\subsection{Proof of Proposition~\ref{prop:overcome_shrinking}}
We here ignore the superscript $\mathcal{T},\ell$ for brevity. By maximizing the overall average norm of $\bar{x}_\mathrm{norm}$, we have, 
\begin{equation}
\begin{aligned}
    N\bar{x}_\text{norm} &= N\cdot\frac{1}{N}\sum_{i\in V} \|\bar{\mathbf{x}}_{i}\|_2^2 \\
    &= \sum_{i\in V} \|\mathbf{x}_{i}-\frac{1}{N}\sum_{i\in V}\mathbf{x}_i\|_2^2 \\
    &=\sum_{i\in V} \left(\|\mathbf{x}_{i}\|_2^2 + \|\frac{1}{V}\sum_{i\in V}\mathbf{x}_i\|_2^2 - \frac{2}{N}\mathbf{x}_i\sum_{i\in V}\mathbf{x}_i\right) \\
    &=\sum_{i\in V}\|\mathbf{x}_i\|_2^2 + \|\mathbf{1}^T\mathbf{x}\|_2^2-2\|\mathbf{1}^T\mathbf{x}\|_2^2 \\
    &=\sum_{i\in V}\mathbf{x}_i^T\mathbf{x}_i + \mathbf{1}^T\mathbf{x}\mathbf{x}^T\mathbf{1}- 2\mathbf{1}^T\mathbf{x}\mathbf{x}^T\mathbf{1} \\
    &=\frac{1}{2}\sum_{i,j\in V}(\mathbf{x}_i^T\mathbf{x}_i + \mathbf{x}_j^T\mathbf{x}_j)-\sum_{i,j\in V}\mathbf{x}_i^T\mathbf{x}_j \\
    & = \frac{1}{2}\sum_{i,j\in V}(\mathbf{x}_i-\mathbf{x}_j)^T(\mathbf{x}_i-\mathbf{x}_j) \\
    &= \frac{1}{2}\sum_{i,j\in V}\|\mathbf{x}_i-\mathbf{x}_j\|_2^2,
\end{aligned}
\label{equ: derivation}
\end{equation}
then we know $\bar{x}_\text{norm}=1/N\cdot\frac{1}{2}\sum_{i,j\in V}\|\mathbf{x}_i-\mathbf{x}_j\|_2^2$, which states that the distance over all node pairs will be enlarged with the coefficient $1/N$. The hyper-parameter before $L_\mathrm{reg}$ will also give some emphasis on the distance over all node pairs.
\end{document}